\documentclass[aps,prb,twocolumn,amsmath,amssymb,superscriptaddress]{revtex4}
\usepackage{bm}
\usepackage{graphicx}

\DeclareMathOperator{\sgn}{sgn}
\DeclareMathOperator{\Img}{\mathrm{Im}}

\DeclareMathOperator{\Sp}{\mathrm{Sp}}

\begin{document}

\title{Absence of "fractional ac Josephson effect" in superconducting junctions}
\author{Mikhail S. Kalenkov}
\affiliation{I.E. Tamm Department of Theoretical Physics, P.N. Lebedev Physical Institute, 119991 Moscow, Russia}
\author{Andrei D. Zaikin}
\affiliation{I.E. Tamm Department of Theoretical Physics, P.N. Lebedev Physical Institute, 119991 Moscow, Russia}
\affiliation{National Research University Higher School of Economics, 101000 Moscow, Russia}
\date{\today}
\begin{abstract}
We develop a microscopic theory of ac Josephson effect in superconducting junctions described by an arbitrary scattering matrix that may include magnetic effects. In the limit of constant in time bias voltage $V$ applied to the junction we derive a formally exact current-phase relation (CPR) that is manifestly $2\pi$-periodic in the Josephson phase $\varphi$ in full accordance with general principles. This our result unambiguously argues against the idea of the so-called "fractional ac Josephson effect" admitting $4\pi$-periodic in $\varphi$ CPR.  We also demonstrate that at any non-zero $V$ quantum dynamics of Andreev bound states becomes non-Hermitian which signals their instability, thus making any  'quasi-equilibrium' description of ac Josephson effect unreliable. We specifically address the limit of highly transparent junctions with magnetic scattering where -- along with super- and excess current terms -- at small $V$ we also recover a non-trivial  $2\pi$-periodic dissipative current with the amplitude $\propto |V|^{1/3}$.
\end{abstract}
\maketitle

\section{Introduction}
In 1962 Josephson formulated his famous equations \cite{Jos}
\begin{gather}
\label{1}
I_s=I_c\sin \varphi,\\
d\varphi/dt=2eV
\label{2}
\end{gather}
describing dynamics of the supercurrent $I_s$ flowing across a tunnel barrier between two superconductors with the phase difference $\varphi (t)$ biased by an external voltage $V$. Here and below the Planck's constant $\hbar$ is set equal to unity, $I_c$ denotes the Josephson critical current and $e$ is the electron charge.

The first of the above equations demonstrates that $I_s$ is a $2\pi$-periodic function of the phase difference $\varphi$ which is a fundamental consequence of the fact that the supercurrent across the Josephson junction is transferred by Cooper pairs with charge $2e$. A detailed discussion of the symmetries for both phase and charge variables describing Josephson junctions can be found, e.g., in the review article \cite{SZ90}.

A large variety of superconducting junctions and weak links has been studied \cite{KGI} which equilibrium current-phase relations (CPR) may deviate from a simple $\sin \varphi$-form  (\ref{1}) but -- in full agreement with the above arguments -- always remain $2\pi$-periodic in $\varphi$ and, hence, can generally be expressed in terms of the Fourier series
\begin{equation}
I_s(\varphi )=\sum_n^{\infty}I_n\sin n\varphi .
\label{3}
\end{equation}

More recently, it was suggested that in some special cases under a small voltage bias CPR of superconducting junctions may actually turn $4\pi$-periodic in $\varphi$. This assertion -- apparently contradicting to the symmetry of the charge states in superconducting junctions  \cite{SZ90} -- was formulated by Kwon et al. \cite{Yak} in the case of junctions formed by $p$-wave superconductors and then extended by Michelsen et al. \cite{Chalmers} to superconducting point contacts with magnetic scattering and by Fu and Kane \cite{FK} to junctions involving topological insulators. 

Let us note that for sufficiently short (with longitudinal size smaller that superconducting coherence length) and not too strongly asymmetric junctions the {\it equilibrium} Josephson current (\ref{3}) can also be recovered with the aid of a simple formula
\begin{equation}
I_s=e\sum_{m=1}^{\mathcal N}\sum_{i}\frac{\partial E^A_{m,i}}{\partial \varphi}f_{m,i},
\label{dEdf}
\end{equation}
where  $E^A_{m,i} (\varphi)$ represents the energy of subgap Andreev levels in the $m$-th conducting channel of the junction and ${\mathcal N}$ is the total number of such channels. It is assumed that every conducting channel hosts Andreev bound states with both spin directions and the index $i$ enummerates the states within each conducting channel. Here $f_{m,i}$  are the filling factors for Andreev bound states. While in equilibrium these filling factors obviously coincide with the Fermi function $f_{m,i}\equiv f_F=1/[1+\exp (E^A_{m,i}/T)]$, thus leaving CPR \eqref{dEdf} $2\pi$-periodic, an attempt to extend this formula to non-equilibrium situations may indeed yield $4\pi$-periodic CPR. The corresponding line of reasoning \cite{Yak,Chalmers,FK} was approximately as follows.

Applying a small bias voltage $V$ one adiabatically drives the system along the set of subgap Andreev levels by 'slowly' changing the phase $\varphi$ in time. Provided these discrete levels touch the continuum at energies equal to the superconducting gap $\pm \Delta$ every time $\varphi$ changes by $2\pi$, the (non-equilibrium) level population $f_{m,i}$ gets reset at these points and, hence, CPR determined from Eq. \eqref{dEdf} still remains $2\pi$-periodic. If, however, discrete levels never touch the continuum, their filling factors $f_{m,i}$ should remain conserved implying that the periodicity of CPR defined by Eq. \eqref{dEdf} is identical to that of Andreev levels. Within a certain parameter range the latter situation is realized, e.g., in superconducting junctions involving spin-active scatterers \cite{Chalmers} where Andreev levels turn out to be $4\pi$-periodic in $\varphi$. Likewise, superconducting junctions hosting Majorana-like modes with energies $E^M_{i}(\varphi) \propto \pm\cos (\varphi/2)$  and $|E^M_{i}|<\Delta$ have been considered \cite{Yak,FK}. In all these cases $4\pi$-periodic CPR follows immediately from Eq. \eqref{dEdf} at non-zero values of $V$. 

Following this scenario one would expect to observe the supercurrent oscillations with frequency equal to $eV$, i.e. to a half of the standard Josephson frequency $2eV$. Such "fractional ac Josephson effect" could then be used to experimentally verify, e.g., the presence of Majorana-like bound states in junctions involving topological insulators by detecting the corresponding current resonances under the influence of external microwave radiation. In experiments with HgTe- and BiSb-based superconducting junctions \cite{M16,Mol,Gre} such resonances (the so-called Shapiro steps) at ('fractional') frequencies $\omega =eV$ have indeed been observed along with "missing" integer Shapiro steps at  $\omega =2eV$. According, e.g., to phenomenological analysis  \cite{4pisteps} these observations could be interpreted in favor of "$4\pi$-periodic Josephson effect" in superconducting junctions under consideration.

Later on the above scenario of "fractional ac Josephson effect" was questioned both theoretically \cite{GZ21} and experimentally \cite{MS}. Galaktionov and one of the present authors \cite{GZ21} argued on general grounds that (i) $4\pi$-periodic CPR would inevitably imply transferring the supercurrent by single electrons with charge $e$ which is hardly possible \cite{FN,FNR} and, on top of that, (ii) the analysis based on Eq. \eqref{dEdf} essentially ignores the mechanism of multiple Andreev reflection (MAR) \cite{MAR} playing an important role in superconducting weak links at non-zero bias voltages. In addition, the overall pattern of Shapiro steps in topologically trivial highly transparent superconducting junctions was found \cite{GZ21} to be similar to that observed in topological Josephson junctions \cite{M16,Mol,Gre}. Furthermore, well-pronounced fractional Shapiro steps along with "missing" integer ones have also been detected in topologically trivial Josephson junctions based on InAs quantum wells \cite{MS}. Thus, the results \cite{GZ21,MS} indicate that caution is needed while interpreting the observations \cite{M16,Mol,Gre} in terms of "$4\pi$-periodic Josephson effect".  

Here we will set up a rigorous calculation which unambiguously argues against the idea of "fractional ac Josephson effect" in superconducting junctions. The structure of the paper is as follows. In Sec. II we define our general model and outline the main formalism to be employed in our calculation. Quasiclassical Green functions for our system are evaluated in Sec. III. Sec. IV is devoted to a microscopic calculation of ac current across our superconducting junction. An important example of spin-active superconducting weak links is considered in Sec. V which is followed by a general discussion of our results in Sec. VI. Some technical details of our calculation are relegated to Appendices.

\section{The model and basic formalism}

Below we will consider a general and rather standard model for a superconducting junction: Two massive superconducting electrodes 
characterized by the order parameter $\Delta_{1,2} = |\Delta_{1,2}| e^{i\chi_{1,2}}$ are separated by an arbitrary normal scatterer of cross-section ${\mathcal A}$ which length is shorter than the superconducting coherence length. The corresponding scattering matrix that accounts for electron transfer across this scatterer is assumed energy independent and can include magnetic effects. Electron transport in superconducting leads will be described within the quasiclassical theory of superconductivity. The corresponding Green-Eilenberger functions evaluated in both superconductors will then be matched by means of appropriate boundary for the normal scatterer connecting the electrodes. 

We will employ the standard Eilenberger equations combined with the Keldysh technique. These equations read \cite{ZRev}
\begin{equation}
-i\bm{v}_F
\nabla \check g 
=
\left[
\varepsilon \hat \tau_3 + e V
+
\check \Delta
,
\check g 
\right].
\label{eilen}
\end{equation}
Here square brackets denote the commutator, $\bm{v}_F$ is the Fermi velocity vector, $V=V(\bm{r},t)$ is the scalar potential, $\hat \tau_3$ is the Pauli matrix and 
\begin{equation}
\check \Delta
=
\begin{pmatrix}
\hat \Delta & 0 \\
0 & \hat \Delta
\end{pmatrix},
\quad
\hat \Delta
=
\begin{pmatrix}
0 & \Delta \\
- \Delta^* & 0
\end{pmatrix}
\end{equation}
with $\Delta$ being the superconducting order parameter. 

In Eq. \eqref{eilen} and below the product of the functions with omitted time arguments should be treated as a convolution. In particular, for the functions in the Wigner representation one has 
\begin{equation}
(AB)(\varepsilon,t)
=
e^{i(\partial_{\varepsilon_a} \partial_{t_b} - \partial_{\varepsilon_b} \partial_{t_a} )/2}
A(\varepsilon_a , t_a) 
B(\varepsilon_b, t_b) 
\Biggr|_{\substack{t_a=t_b=t, \\ \varepsilon_a = \varepsilon_b = \varepsilon}},
\end{equation}
which is equivalent to 
\begin{equation}
(AB)(t,t') = 
\int A(t, \tilde t) B(\tilde t, t') d \tilde t
\end{equation}
in the conventional representation depending on the two time arguments $t$ and $t'$. Likewise, the term $\varepsilon$ in the Wigner representation in the right hand-side of Eq. \eqref{eilen} corresponds to the differential operator 
\begin{equation}
\varepsilon \Leftrightarrow i \delta'(t-t')
\end{equation}
that depends on both $t$ and $t'$. Hence, the time derivatives are effectively contained in the term $\varepsilon$.

The Green-Eisenberger functions $\check g$ are $8\times 8$ matrices in the Spin$\otimes$Nambu$\otimes$Keldysh space  
\begin{equation}
\check g
=
\begin{pmatrix}
\hat g^R & \hat g^K \\
0 & \hat g^A
\end{pmatrix},
\end{equation}
consisting of retarded $\hat g^R$, advanced $\hat g^A$ and Keldysh $\hat g^K$ $4\times 4$ matrix functions.  
Electric current density $\bm{j}$ in our structure is evaluated by means of the formula
\begin{gather}
\bm{j} =  \dfrac{eN_0}{8}
\int
d \varepsilon
\left<\bm{v}_F \Sp [\hat \tau_3\hat g^K (\varepsilon, t)]\right>,
\label{cure}
\end{gather}
where $\hat g^K (\varepsilon, t)$ is the Keldysh-Green-Eilenberger function in the Wigner representation, $N_0$ is the density of states at the Fermi energy per spin direction and angular brackets denote averaging over directions of the Fermi velocity vector.

The normal scatterer in-between two superconducting electrodes is described by the scattering matrices
\begin{gather}
\mathcal{S}=
\begin{pmatrix}
S_{11} & S_{12} \\
S_{21} & S_{22}
\end{pmatrix}, \quad 
\underline{\mathcal{S}}=
\begin{pmatrix}
\underline{S}_{11} & \underline{S}_{12} \\
\underline{S}_{21} & \underline{S}_{22}
\end{pmatrix}
\end{gather}
respectively for electron-like and hole-like excitations. It can also be convenient to combine these two matrices into a single one 
\begin{equation}
\mathbb{S}_n = 
\begin{pmatrix}
\mathcal{S}^+ &  0\\
0 & \underline{\mathcal{S}}
\end{pmatrix},
\end{equation}
which will be employed below in our calculation.

With the aid of these matrices one can formulate the boundary conditions for the Green functions of both superconductors. In the case of non-magnetic scatterers these boundary conditions were derived by Zaitsev \cite{Zaitsev84} and then generalized by Millis et al. \cite{Millis88} to spin-active interfaces. The resulting boundary conditions have the form of nonlinear matrix equations for quasiclassical Green functions on both sides of the interface and are rather complicated to deal with. Certain simplifications could be achieved employing the so-called Riccati parameterization of the Green functions \cite{Eschrig00, Zhao04}. Below we will essentially follow the notations adopted in these papers.

For the retarded and advanced Green functions we set  
\begin{gather}
\hat g^R=
\hat N^R
\begin{pmatrix}
1+\gamma^R \tilde \Gamma^R & 2\gamma^R \\
-2 \tilde \Gamma^R & -1- \tilde \Gamma^R  \gamma^R \\
\end{pmatrix},
\label{graparamR}
\\
\hat g^A=
-
\hat N^A
\begin{pmatrix}
1+\Gamma^A \tilde \gamma^A & 2\Gamma^A \\
-2 \tilde \gamma^A & -1- \tilde \gamma^A  \Gamma^A \\
\end{pmatrix},
\label{graparamA}
\end{gather}
where $\hat N^A$ are diagonal matrices
\begin{gather}
\hat N^R=
\begin{pmatrix}
(1-\gamma^R \tilde \Gamma^R)^{-1} & 0 \\
0 & (1-\tilde \Gamma^R  \gamma^R )^{-1} \\
\end{pmatrix},
\label{nrparamR}
\\
\hat N^A=
\begin{pmatrix}
(1-\Gamma^A \tilde \gamma^A)^{-1} & 0 \\
0 & (1-\tilde \gamma^A  \Gamma^A )^{-1} \\
\end{pmatrix}.
\label{nrparamA}
\end{gather}
Keldysh Green functions are parameterized by means of the same Riccati amplitudes and the two distribution functions $x$ and $\tilde X$
\begin{multline}
\hat g^K=
2
\hat N^R
\\\times
\begin{pmatrix}
x^K - \gamma^R  \tilde X^K  \tilde \Gamma^A &
-\gamma^R  \tilde X^K + x^K  \Gamma^A \\
-\tilde \Gamma^R  x^K + \tilde X^K  \tilde \Gamma^A &
\tilde X^K - \tilde \Gamma^R  x^K  \Gamma^A \\
\end{pmatrix}
\hat N^A.
\label{gkparam}
\end{multline}
\begin{figure}
\includegraphics[width=80mm]{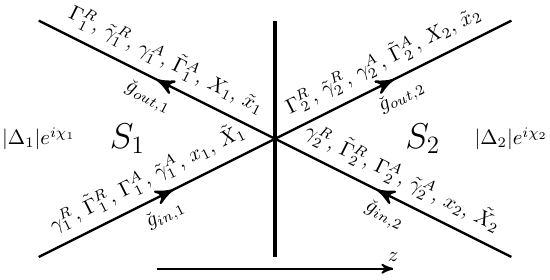}
\caption{Graphical representation of the boundary conditions setting the relations between the Green functions for both incoming and outgoing velocity directions on the two sides of the interface. Each Green function is parameterized by an appropriate set of Riccati amplitudes and distribution functions.}
\label{sis}
\end{figure}
These notations will be chosen for the Green functions $\hat g^{R,A,K}_{in,k}$ with velocity directions incoming to the interface between two superconducting electrodes labeled by the index $k=1,2$. The Green functions $\hat g^{R,A,K}_{out,k}$ with outgoing velocity directions are parametrized in the same way except capital letters $\Gamma$ and $X$ are replaced by the lowercase ones $\gamma$ and $x$ and vice versa. 

Capital Riccati amplitudes $\Gamma$ and the distribution functions $X$ can be expressed in terms of the corresponding lowercase functions $\gamma$ and $x$. For Riccati amplitudes one has \cite{Eschrig00, Zhao04}
\begin{gather}
\Gamma_1^R = r_{1l}^R \gamma_1^R \underline{S}_{11}^+  + t_{1l}^R \gamma_2^R \underline{S}_{12}^+,
\label{Gamma1R}
\\
\tilde \Gamma_1^R = \tilde r_{1l}^R \tilde \gamma_1^R S_{11}  + \tilde t_{1l}^R \tilde \gamma_2^R S_{21},
\label{tGamma1R}
\\
\Gamma_1^A = S^+_{11} \gamma_1^A r_{1r}^A  + S^+_{21} \gamma_2^A t_{1r}^A,
\label{Gamma1A}
\\
\tilde \Gamma_1^A = \underline{S}_{11} \tilde \gamma_1^A \tilde r_{1r}^A  + \underline{S}_{12} \tilde \gamma_2^A \tilde t_{1r}^A,
\label{tGamma1A}
\end{gather}
where $r_{1l}^{R,A}$ and $t_{1l}^{R,A}$ are effective reflection and transmission amplitudes
\begin{gather}
r_{1l}^R = [(\beta_{21}^R)^{-1} S_{11}^+ - (\beta_{22}^R)^{-1} S_{12}^+ ]^{-1} (\beta_{21}^R)^{-1},
\\
t_{1l}^R = -[(\beta_{21}^R)^{-1} S_{11}^+ - (\beta_{22}^R)^{-1} S_{12}^+ ]^{-1} (\beta_{22}^R)^{-1},
\\
\tilde r_{1l}^R = [(\tilde \beta_{21}^R)^{-1} \underline{S}_{11} - (\tilde \beta_{22}^R)^{-1} \underline{S}_{21} ]^{-1} (\tilde \beta_{21}^R)^{-1},
\\
\tilde t_{1l}^R = -[(\tilde \beta_{21}^R)^{-1} \underline{S}_{11} - (\tilde \beta_{22}^R)^{-1} \underline{S}_{21} ]^{-1} (\tilde \beta_{22}^R)^{-1},
\\
r_{1r}^A = (\tilde \beta_{21}^A)^{-1} [\underline{S}^+_{11} (\tilde \beta_{21}^A)^{-1} - \underline{S}^+_{21} (\tilde \beta_{22}^A)^{-1} ]^{-1} ,
\\
t_{1r}^A = -(\tilde \beta_{22}^A)^{-1} [\underline{S}^+_{11} (\tilde \beta_{21}^A)^{-1} - \underline{S}^+_{21} (\tilde \beta_{22}^A)^{-1} ]^{-1} ,
\\
\tilde r_{1r}^A = (\beta_{21}^A)^{-1} [S_{11} (\beta_{21}^A)^{-1} - S_{12} (\beta_{22}^A)^{-1} ]^{-1} ,
\\
\tilde t_{1r}^A = -(\beta_{22}^A)^{-1} [S_{11} (\beta_{21}^A)^{-1} - S_{12} (\beta_{22}^A)^{-1} ]^{-1} ,
\end{gather}
where the functions $\beta^{R,A}$ and $\tilde \beta^{R,A}$ are defined as
\begin{gather}
\beta^R_{ij} = S_{ij}^+ - \gamma_j^R \underline{S}_{ij}^+ \tilde \gamma_i^R,
\quad
\tilde \beta^R_{ij} = \underline{S}_{ji} - \tilde \gamma_j^R S_{ji} \gamma_i^R,
\label{betaR}
\\
\beta^A_{ij} = S_{ij} - \gamma_i^A \underline{S}_{ij} \tilde \gamma_j^A,
\quad
\tilde \beta^A_{ij} = \underline{S}_{ji}^+ - \tilde \gamma_i^A S_{ji}^+ \gamma_j^A,
\label{betaA}
\end{gather}

Boundary conditions for the distribution functions read
\begin{gather}
X_1 = r_{1l}^R x_1 \tilde r_{1r}^A + t_{1l}^R x_2 \tilde t_{1r}^A - a_{1l}^R \tilde x_2 \tilde a_{1r}^A,
\label{Xbound}
\\
\tilde X_1 = \tilde r_{1l}^R \tilde x_1 r_{1r}^A + \tilde t_{1l}^R \tilde x_2 t_{1r}^A - \tilde a_{1l}^R x_2 a_{1r}^A,
\label{Xboundtild}
\end{gather}
They contain the so-called branch conversion amplitudes 
\begin{gather}
a_{1l}^R = (\Gamma_1^R \underline{S}_{11} - S_{11} \gamma^R_1) (\tilde \beta^R_{12})^{-1},
\label{a1lR}
\\
\tilde a_{1r}^A = (\tilde \beta^A_{12})^{-1} ( \underline{S}_{11}^+ \tilde \Gamma_1^A - \tilde \gamma^A_1 S_{11}^+ ),
\label{ta1rA}
\\
\tilde a_{1l}^R = (\tilde \Gamma_1^R S_{11}^+ - \underline{S}_{11}^+ \tilde \gamma^R_1) (\beta^R_{12})^{-1},
\label{ta1lR}
\\
a_{1r}^A = (\beta^A_{12})^{-1} ( S_{11} \Gamma_1^A - \gamma^A_1 \underline{S}_{11} ).
\label{a1rA}
\end{gather}
In Eqs. \eqref{Gamma1R}-\eqref{a1rA} we follow the notations introduced in Ref. \onlinecite{Zhao04}. An explicit meaning of these notations will be clarified in the next section. 

\section{Green functions}
Let us combine Riccati amplitudes and the distribution functions at the opposite sides of the interface by defining the following matrices in the terminal space
\begin{multline}
\hat p = 
\begin{pmatrix}
p_1 & 0 \\
0 & p_2
\end{pmatrix},
\\
p = \gamma^{R,A}, \ \tilde \gamma^{R,A}, \Gamma^{R,A}, \ \tilde \Gamma^{R,A}, x,\ \tilde x, \ X, \ \tilde X.
\end{multline}
Similarly for the branch conversion amplitudes we define
\begin{gather}
\hat a^R =
\begin{pmatrix}
0 & a_{1l}^R \\
a_{2l}^R & 0
\end{pmatrix},
\quad
\hat{\tilde a}^R =
\begin{pmatrix}
0 & \tilde a_{1l}^R \\
\tilde a_{2l}^R & 0
\end{pmatrix},
\\
\hat a^A =
\begin{pmatrix}
0 & a_{2r}^A \\
a_{1r}^A & 0
\end{pmatrix},
\quad
\hat{\tilde a}^A =
\begin{pmatrix}
0 & \tilde a_{2r}^A \\
\tilde a_{1r}^A & 0
\end{pmatrix}.
\end{gather} 

Making use of the above relations and performing simple matrix transformations one can rewrite the boundary conditions for Riccati amplitudes in a compact matrix form. In order to evaluate the retarded Green function it is necessary to find the combinations $(1 - \hat \Gamma^R \hat{\tilde \gamma}^R)^{-1}$ and $(1 - \hat{\tilde \Gamma}^R \hat \gamma^R)^{-1}$. It is straightforward to verify that the following two matrix equations
\begin{gather}
(1 - \hat \Gamma^R \hat{\tilde \gamma}^R)^{-1}
(1-\hat a^R \hat{\tilde \gamma}^R)
=
(
\mathcal{S}^+
-
\hat \gamma^R
\underline{\mathcal{S}}^+
\hat{\tilde \gamma}^R
)^{-1} \mathcal{S}^+,
\label{Gamma1}
\\
(1 - \hat{\tilde \Gamma}^R \hat \gamma^R)^{-1}
(1 - \hat{\tilde a}^R \hat \gamma^R)
=
(\underline{\mathcal{S}} -
\hat{\tilde \gamma}^R
\mathcal{S}
\hat \gamma^R)^{-1} \underline{\mathcal{S}}
\label{Gamma2}
\end{gather}
are fully equivalent to the boundary conditions \eqref{Gamma1R} and \eqref{tGamma1R}. Similar equations for advanced Riccati amplitudes $\hat \Gamma^A$ and $\hat{\tilde \Gamma}^A$ can be recovered from Eqs. \eqref{Gamma1}, \eqref{Gamma2} combined with the symmetry relations 
\begin{equation}
\hat y^A = (\hat{\tilde y}^R)^+, 
\quad
\hat{\tilde y}^A = (\hat y^R)^+,
\quad  y=\gamma,\ \Gamma,\ a.
\end{equation}
A complete set of matrix equations involving different Riccati amplitudes is collected in Appendix \ref{Aric}.

Equations \eqref{Gamma1} and  \eqref{Gamma2} -- being compact and transparent -- provide a useful alternative for the boundary conditions \eqref{Gamma1R} and \eqref{tGamma1R}. For instance, the boundary conditions \eqref{Gamma1R} can be immediately recovered evaluating the ``11'' block in the terminal space of Eq. \eqref{Gamma1} with the aid of the matrix block inversion formula \eqref{blockinv} (see Appendix B). Making use of the standard matrix transformation one can also verify the equivalence between Eqs. \eqref{Gamma1} and \eqref{Gamma7}. Evaluating the ``11'' block in Eq. \eqref{Gamma7} in the same manner, we obtain an equivalent representation of Eq. \eqref{Gamma1R}
\begin{equation}
\Gamma_1^R = S_{11} \gamma_1^R r_{1r}^R + S_{12}\gamma_2^R t_{2r}^R,
\label{Gamma1R2}
\end{equation}
where $\gamma$, $S$, $r$, and $t$ elements enter in a reverse order, and the coefficients $r_{1r}^R$ and $t_{2r}^R$ are defined as 
\begin{gather}
r_{1r}^R
=(\tilde\beta_{12}^R)^{-1}
[\underline{S}_{11} (\tilde\beta_{12}^R)^{-1} - \underline{S}_{12} (\tilde\beta_{22}^R)^{-1}],
\\
t_{1r}^R
=-(\tilde\beta_{22}^R)^{-1}
[\underline{S}_{11} (\tilde\beta_{12}^R)^{-1} - \underline{S}_{12} (\tilde\beta_{22}^R)^{-1}].
\end{gather}
This definition is somewhat different from that for $r_{1l}^R$ and $t_{2l}^R$.
An equivalent representation for other capital Riccati amplitudes can be established analogously.

With the aid of the above boundary conditions one can derive explicit expressions for retarded Green functions at the interface. They read
\begin{multline}
\hat g^R_{in} 
=
2
\begin{pmatrix}
\mathcal{S}^+ & 0 \\
0 & 1
\end{pmatrix}
\begin{pmatrix}
\mathcal{S}^+ & \hat \gamma^R \\
\hat{\tilde \gamma}^R & \underline{\mathcal{S}}
\end{pmatrix}^{-1}
\begin{pmatrix}
1 & 0 \\
0 & - \underline{\mathcal{S}}
\end{pmatrix}
-
\hat \tau_3
\\-
2
\begin{pmatrix}
- \hat \gamma^R \\ 1
\end{pmatrix}
(1 - \hat{\tilde \Gamma}^R \hat \gamma^R)^{-1} \hat{\tilde a}^R
\begin{pmatrix}
1 & \hat \gamma^R
\end{pmatrix},
\label{gRin}
\end{multline}
and
\begin{multline}
\hat g^R_{out} 
=
-\hat \tau_3
+
2
\begin{pmatrix}
1 & 0 \\
0 & \underline{\mathcal{S}}
\end{pmatrix}
\begin{pmatrix}
\mathcal{S}^+ & \hat \gamma^R  \\
\hat{\tilde \gamma}^R & \underline{\mathcal{S}}
\end{pmatrix}^{-1}
\begin{pmatrix}
\mathcal{S}^+ & 0 \\
0 & -1
\end{pmatrix}
\\+
2
\begin{pmatrix}
1 \\ -\hat{\tilde \gamma}^R
\end{pmatrix}
(1-\hat \Gamma^R \hat{\tilde \gamma}^R)^{-1}
\hat a^R
\begin{pmatrix}
\hat{\tilde \gamma}^R & 1
\end{pmatrix},
\label{gRout}
\end{multline}
where we introduced the matrices 
\begin{gather}
\hat g^{R,A}_{in}
=
\begin{pmatrix}
\hat g^{R,A}_{in,1} & 0 \\ 0 &  \hat g^{R,A}_{in,2}.
\end{pmatrix},
\
\hat g^{R,A}_{out}
=
\begin{pmatrix}
\hat g^{R,A}_{out,1} & 0 \\ 0 &  \hat g^{R,A}_{out,2}.
\end{pmatrix}
\end{gather}
combining the corresponding Green functions in both terminals. 
It is important to point out that the matrices $\hat g^R_{in} $ and $\hat g^R_{out} $ are diagonal in the terminal space, whereas the last terms in Eqs. \eqref{gRin} and \eqref{gRout} contain only off-diagonal elements in this space. Hence, in order to evaluate the Green functions it suffices to keep only the diagonal elements in the first two terms in the right-hand side of these equations. 

Advanced Green functions can then easily be recovered from the identities 
\begin{gather}
\nonumber
\hat g^A =  -\hat \tau_3 (\hat g^R)^+ \hat \tau_3,
\end{gather}
whereas Keldysh Green functions can be explicitly evaluated with the aid of the boundary conditions for the distribution functions \eqref{Xbound} and \eqref{Xboundtild} combined with the identities 
\begin{gather}
(1 - \hat \Gamma^R \hat{\tilde \gamma}^R)^{-1}
\begin{pmatrix}
r_{1l}^R & t_{1l}^R \\
t_{2l}^R & r_{2l}^R
\end{pmatrix}
=
(
\mathcal{S}^+
-
\hat \gamma^R
\underline{\mathcal{S}}^+
\hat{\tilde \gamma}^R
)^{-1},
\\
(1 - \hat{\tilde \Gamma}^R \hat \gamma^R)^{-1}
\begin{pmatrix}
\tilde r_{1l}^R & \tilde t_{1l}^R \\
\tilde t_{2l}^R & \tilde r_{2l}^R
\end{pmatrix}
=
(\underline{\mathcal{S}} -
\hat{\tilde \gamma}^R
\mathcal{S}
\hat \gamma^R)^{-1},
\\
\begin{pmatrix}
r_{1r}^A & t_{2r}^A \\
t_{1r}^A & r_{2r}^A
\end{pmatrix}
(1 - \hat{\tilde \gamma}^A \hat \Gamma^A)^{-1}
=
(
\underline{\mathcal{S}}^+
-
\hat{\tilde \gamma}^A
\mathcal{S}^+
\hat\gamma^A )^{-1},
\\
\begin{pmatrix}
\tilde r_{1r}^A & \tilde t_{2r}^A \\
\tilde t_{1r}^A & \tilde r_{2r}^A
\end{pmatrix}
(1 - \hat\gamma^A
\hat{\tilde \Gamma}^A)^{-1}
=
(\mathcal{S}
-
\hat \gamma^A
\underline{\mathcal{S}}
\hat{\tilde \gamma}^A  )^{-1}.
\end{gather}

Alternatively, the whole Keldysh structure of the Green function matrix can be reconstructed from the expressions for the retarded Green function \eqref{gRin} and \eqref{gRout} by means of a formal replacement 
\begin{gather}
\hat \gamma^R \Rightarrow
\begin{pmatrix}
\hat \gamma^R & -\hat x^K (\hat{\tilde\gamma}^A)^{-1} \\
0 & (\hat{\tilde\gamma}^A)^{-1} \\
\end{pmatrix},
\quad
\hat{\tilde \gamma}^R \Rightarrow
\begin{pmatrix}
\hat{\tilde\gamma}^R & \hat{\tilde x}^K (\hat \gamma^A)^{-1} \\
0 & (\hat \gamma^A)^{-1} \\
\end{pmatrix},
\\
\hat \Gamma^R \Rightarrow
\begin{pmatrix}
\hat \Gamma^R & -\hat X^K (\hat{\tilde\Gamma}^A)^{-1} \\
0 & (\hat{\tilde\Gamma}^A)^{-1} \\
\end{pmatrix},
\quad
\hat{\tilde \Gamma}^R \Rightarrow
\begin{pmatrix}
\hat{\tilde\Gamma}^R & \hat{\tilde X}^K (\hat\Gamma^A)^{-1} \\
0 & (\hat\Gamma^A)^{-1} \\
\end{pmatrix}.
\end{gather}
As a result, we obtain 
\begin{widetext}
\begin{multline}
\check g_{in}
=
\begin{pmatrix}
\hat g^R_{in} & \hat f^R_{in} & \hat g^K_{in} & \hat f^K_{in}\\
\hat {\tilde f}^R_{in} & \hat{\tilde g}^R_{in} & \hat {\tilde f}^K_{in} & \hat{\tilde g}^K_{in}\\
0 & 0 & \hat g^A_{in} & \hat f^A_{in} \\
0 & 0 & \hat {\tilde f}^A_{in} & \hat{\tilde g}^A_{in}
\end{pmatrix}
=-
\begin{pmatrix}
1 & 0 & 0 & 0 \\
0 & -1 & 0 & 0 \\
0 & 0 & -1 & 0 \\
0 & 0 & 0 & 1 \\
\end{pmatrix}
+
2
\begin{pmatrix}
\mathcal{S}^+ & 0 & 0 & 0\\
0 & 1 & 0 & 0\\
0 & 0 & 1 & 0 \\
0 & 0 & 0 & \underline{\mathcal{S}}^+
\end{pmatrix}
\begin{pmatrix}
\mathcal{S}^+  &  \gamma^R  & x^K & 0 \\
\tilde \gamma^R   & \underline{\mathcal{S}} & 0 & -\tilde x^K\\
0 & 0 & \mathcal{S} & \gamma^A   \\
0 & 0 & \tilde\gamma^A & \underline{\mathcal{S}}^+
\end{pmatrix}^{-1}
\begin{pmatrix}
1 & 0  & 0 & 0\\
0 & -\underline{\mathcal{S}} & 0 & 0 \\
0 & 0 & -\mathcal{S} & 0 \\
0 & 0 & 0 & 1
\end{pmatrix}
+\\+
(\text{off-diagonal in terminal space terms}),
\label{gKin}
\end{multline}
and 
\begin{multline}
\check g_{out}
=
\begin{pmatrix}
\hat g^R_{out} & \hat f^R_{out} & \hat g^K_{out} & \hat f^K_{out}\\
\hat {\tilde f}^R_{out} & \hat{\tilde g}^R_{out} & \hat {\tilde f}^K_{out} & \hat{\tilde g}^K_{out}\\
0 & 0 & \hat g^A_{out} & \hat f^A_{out} \\
0 & 0 & \hat {\tilde f}^A_{out} & \hat{\tilde g}^A_{out}
\end{pmatrix}
=-
\begin{pmatrix}
1 & 0 & 0 & 0 \\
0 & -1 & 0 & 0 \\
0 & 0 & -1 & 0 \\
0 & 0 & 0 & 1 \\
\end{pmatrix}
+
2
\begin{pmatrix}
1 & 0  & 0 & 0\\
0 & \underline{\mathcal{S}} & 0 & 0 \\
0 & 0 & \mathcal{S} & 0 \\
0 & 0 & 0 & 1
\end{pmatrix}
\begin{pmatrix}
\mathcal{S}^+  &  \gamma^R  & x^K & 0 \\
\tilde \gamma^R   & \underline{\mathcal{S}} & 0 & -\tilde x^K\\
0 & 0 & \mathcal{S} & \gamma^A   \\
0 & 0 & \tilde\gamma^A & \underline{\mathcal{S}}^+
\end{pmatrix}^{-1}
\begin{pmatrix}
\mathcal{S}^+ & 0 & 0 & 0\\
0 & -1 & 0 & 0\\
0 & 0 & -1 & 0 \\
0 & 0 & 0 & \underline{\mathcal{S}}^+
\end{pmatrix}
+\\+
(\text{off-diagonal in terminal space terms}).
\label{gKout}
\end{multline}
\end{widetext}

Eqs. \eqref{gKin} and \eqref{gKout} fully determine retarded, advanced and Keldysh components of the interface Green function matrix in terms of the lowercase Riccati amplitudes and the distribution functions as well as the elements of the scattering matrix. Below we will employ these equations in order to evaluate electric current across our superconducting weak link out of equilibrium. 

\section{Electric current and CPR}

In order to proceed let us define the matrix Andreev amplitude 
\begin{equation}
\Gamma=
\begin{pmatrix}
0  &  \hat\gamma^R\\
\hat{\tilde \gamma}^R   & 0
\end{pmatrix}
\end{equation}
and introduce the distribution functions 
\begin{equation}
\begin{pmatrix}
\hat x & 0 \\
0 & - \hat{\tilde x}
\end{pmatrix}
= H - \Gamma H \Gamma^+,
\end{equation}
where $H$ is the diagonal matrix of the form
\begin{equation}
H=
\begin{pmatrix}
h_1 & 0 & 0 & 0 \\
0 & h_2 & 0 & 0 \\
0 & 0 & \tilde h_1 & 0 \\
0 & 0 & 0 & \tilde h_2 \\
\end{pmatrix}
\end{equation}
with the elements equal to the distribution functions of electron-like and hole-like quasiparticles coming from the bulk of the first and the second superconductors. In equilibrium, the Fourier transformed matrix $H\equiv H_{eq}$ obviously equals to the unity matrix times 
\begin{equation}
h_0(\varepsilon) = \tanh \dfrac{\varepsilon}{2T}.
\label{h0}
\end{equation}

Making use of Eqs. \eqref{cure}, \eqref{gKin} and \eqref{gKout} we can now evaluate electric current $I$ flowing across our Josephson weak link. We obtain
\begin{multline}
I =   \dfrac{\pi eN_0\mathcal{A}}{8}
\bigl<{v}_{F,z}\Theta({v}_{F,z}) \Sp (\hat o_3 \hat \tau_3 [\hat g^K_{in} (t, t) 
-\\-
\hat g^K_{out} (t, t)])\bigr>
=
\dfrac{\pi eN_0\mathcal{A}}{4}
\left<{v}_{F,z}\Theta({v}_{F,z}) \mathcal{T} (t)\right>.
\label{cur}
\end{multline}
where ${v}_{F,z}$ is the component of the Fermi velocity normal to the junction area, $\Theta (v)$ is the Heaviside step function, $\hat o_3 $ is the Pauli matrix in the terminal space, and we defined 
\begin{multline}
\mathcal{T} = 
\Sp \Bigl\{
[\mathbb{S}_n + \Gamma]^{-1}
(H - \Gamma H \Gamma^+)
\\\times
[\mathbb{S}_n^+ + \Gamma^+]^{-1}(\mathbb{S}_n^+
\hat \tau_3 \hat o_3
\mathbb{S}_n - \hat \tau_3 \hat o_3)
\Bigr\}.
\label{T}
\end{multline}
Equations \eqref{cur} and \eqref{T} represent a formally exact result for the current $I$ which has a transparent and appealing matrix structure
convenient for further calculations. 

Let us also note that averaging over the Fermi velocity directions in Eq. \eqref{cur} can be replaced by the sum over the junction conducting
channels (for a given spin direction)
\begin{equation}
N_0{\mathcal A}\left<{v}_{F,z}\Theta({v}_{F,z}) (...)\right> \to \frac{1}{2\pi}\sum_{m=1}^{\mathcal N}(...),
\label{ch}
\end{equation} 
whereas the summation over the spin variable is performed under Sp in Eq. \eqref{T}. 
By setting ${\mathcal N} =1$, we get
\begin{equation}
I=e{\mathcal T}(t)/8.
\label{onech}
\end{equation}
This simple equation directly relates the current $I$ and the combination \eqref{T} in the case of single channel junctions which we will merely address further below. Generalization to an arbitrary number of channels is straightforward and does not require any additional calculation of ${\mathcal T}$.

Let us bias our Josephson junction by the time dependent voltage $V(t)$. Then according to Eq. \eqref{2} the phase difference across the junction acquires the time dependence and reads
\begin{equation}
\varphi (t)=\chi_1(t)-\chi_2(t)= 2e\int^tV(t')dt'.
\label{21}
\end{equation}
Without loss of generality we may set $\chi_1=-\chi_2=\varphi/2$. Then we obtain
\begin{gather}
\Gamma(t,t') = e^{i\varphi(t) \hat \tau_3 \hat o_3/4} \Gamma_{eq}(t-t') e^{-i\varphi(t') \hat \tau_3 \hat o_3/4},
\\
H(t,t') = e^{i\varphi(t) \hat \tau_3 \hat o_3/4} H_{eq}(t-t') e^{-i\varphi(t') \hat \tau_3 \hat o_3/4}.
\end{gather}
With the aid of these relations we can transform Eq. \eqref{T} in the following way
\begin{multline}
{\mathcal T} =
\Sp
\Bigl[
(\tilde{\mathbb{S}}_n + \Gamma_{eq})^{-1} 
(H_{eq} - \Gamma_{eq} H_{eq} \Gamma_{eq}^+)
\\\times
(\tilde{\mathbb{S}}_n^+ + \Gamma_{eq}^+)^{-1}
\left(\tilde{\mathbb{S}}_n^+
\hat o_3 \hat \tau_3
\tilde{\mathbb{S}}_n 
-
\hat o_3 \hat \tau_3
\right) 
\Bigr],
\label{T2}
\end{multline}
where we introduced the time dependent scattering matrix $\tilde{\mathbb{S}}_n$ 
\begin{gather}
\tilde{\mathbb{S}}_n(t) =e^{-i\varphi(t) \hat \tau_3 \hat o_3/4} \mathbb{S}_n 
e^{i\varphi(t) \hat \tau_3 \hat o_3/4}.
\end{gather}
Taking the derivative of the scattering matrix $\tilde{\mathbb{S}}_n$ with respect to the phase $\varphi$,
\begin{equation}
\partial_\varphi \tilde{\mathbb{S}}_{n}(t)
=
\dfrac{i}{4}[- \hat \tau_3 \hat o_3\tilde{\mathbb{S}}_n(t) + \tilde{\mathbb{S}}_n(t) \hat \tau_3 \hat o_3 ],
\end{equation}
we can also rewrite Eq. \eqref{T2} in a more compact form
\begin{multline}
{\mathcal T} =
4i\Sp
\Bigl[
(\tilde{\mathbb{S}}_n + \Gamma_{eq})^{-1} 
(H_{eq} - \Gamma_{eq} H_{eq} \Gamma_{eq}^+)
\\\times
(\tilde{\mathbb{S}}_n^+ + \Gamma_{eq}^+)^{-1}\tilde{\mathbb{S}}_n^+
\partial_\varphi\tilde{\mathbb{S}}_{n}(t)
\Bigr].
\label{T3}
\end{multline}

\subsection{Equilibrium}
Let us first send the bias voltage to zero, $V \to 0$, and verify that in equilibrium our approach yields the standard results for the Josephson current. Provided the phase variable $\varphi$ does not depend on time it is convenient to make use of the Fourier representation. With the aid of the identity
\begin{multline}
(\tilde{\mathbb{S}}_n + \Gamma_{eq})^{-1} 
(1 - \Gamma_{eq} \Gamma_{eq}^+)
(\tilde{\mathbb{S}}_n^+ + \Gamma_{eq}^+)^{-1}
\\=
-
1
+
(\tilde{\mathbb{S}}_n + \Gamma_{eq})^{-1}\tilde{\mathbb{S}}_n
+
\tilde{\mathbb{S}}_n^+ (\tilde{\mathbb{S}}_n^+ + \Gamma_{eq}^+)^{-1}
\end{multline}
we can rewrite Eq. \eqref{T3} in the form 
\begin{equation}
{\mathcal T} =
-
8\int \dfrac{d \varepsilon}{2\pi}
H_{eq}(\varepsilon)
\Img \dfrac{\partial_\varphi\det|\tilde{\mathbb{S}}_n + \Gamma_{eq}(\varepsilon)|}{
\det|\tilde{\mathbb{S}}_n + \Gamma_{eq}(\varepsilon)|}.
\label{Teq}
\end{equation}
Combining this expression with Eq. \eqref{onech} we immediately arrive at the standard 
expression for the supercurrent 
\begin{equation}
I_s= - \frac{e}{2}\sum_{i}
\dfrac{\partial \varepsilon_i}{\partial \varphi}h_0(\varepsilon_i),
\label{IS}
\end{equation}
where the sum runs over all subgap bound states of the problem for both spin directions. Obviously, 
this result is identical to Eq. \eqref{dEdf} for ${\mathcal N} =1$.

\subsection{Constant voltage bias}
Now let us assume that the bias voltage $V$ remains time independent. In this case one obviously has
\begin{equation}
\varphi = 2eVt.
\label{phiL}
\end{equation}

To begin with, we note that provided the phase $\varphi$ \eqref{phiL} changes by $2\pi$ the scattering matrix $\tilde{\mathbb{S}}_n$ gets transformed as 
\begin{equation}
\tilde{\mathbb{S}}_n(t + \pi/(eV)) = \hat o_3 \hat \tau_3 \tilde{\mathbb{S}}_n(t )\hat o_3 \hat \tau_3
=
\hat o_3 \tilde{\mathbb{S}}_n(t )\hat o_3,
\end{equation}
where we made use of the fact that the scattering matrix $\tilde{\mathbb{S}}_n$ is diagonal in the Nambu space. On the other hand, the matrix $\Gamma_{eq}$ is diagonal in the terminal space. Employing the identity
\begin{equation}
\hat o_3 \Gamma_{eq} \hat o_3 = \Gamma_{eq}
\end{equation}
we immediately arrive at the conclusion that ${\mathcal T}(t)$ \eqref{T3} is a periodic function of time with period $\pi/(eV)$. Hence, the current $I$ across our superconducting junction should also depend periodically on time with the same period.

In order to evaluate the general expression for ${\mathcal T}$ \eqref{T} one needs to find the inverse operator $[\mathbb{S}_n + \Gamma]^{-1}$
by resolving the following integral matrix equation
\begin{equation}
{\mathbb{S}}_n  Y(t,t') + 
\int_{-\infty}^t 
\Gamma(t, \tilde t)Y(\tilde t,t') d\tilde t
=
\delta(t-t').
\label{inv}
\end{equation}
This task can conveniently be performed employing a mixed time-frequency representation
\begin{equation}
Y(t, \varepsilon)
=
\int d t' e^{i \varepsilon (t - t')} Y(t, t').
\end{equation}
In this representation expression the function ${\mathcal T}$ \eqref{T} acquires a relatively simple form 
\begin{multline}
{\mathcal T}=
\int 
\dfrac{d \varepsilon}{2\pi}
\Sp
\Bigl\{
Y(t, \varepsilon)
Q(\varepsilon, eV)
\\\times
Y^+(t, \varepsilon)
\left(\mathbb{S}_n^+ \hat o_3 \hat \tau_3 \mathbb{S}_n  
- \hat o_3 \hat \tau_3 \right) 
\Bigr\},
\label{TVconst}
\end{multline}
where we defined 
\begin{multline}
Q(\varepsilon, eV)
=
[1 - \Gamma_{eq}(\varepsilon + eV/2)\Gamma_{eq}^+(\varepsilon + eV/2) ] h_0(\varepsilon + eV/2) \hat P_+
\\+
[1 - \Gamma_{eq}(\varepsilon - eV/2)\Gamma_{eq}^+(\varepsilon - eV/2)] h_0(\varepsilon - eV/2) \hat P_-,
\end{multline}
and introduced the matrices 
\begin{equation}
\hat P_{\pm}
=
\dfrac{1}{2}
(1 \pm \hat o_3 \hat \tau_3).
\end{equation}
 
For the function $Y (t, \varepsilon)$ we get 
\begin{multline}
Y (t, \varepsilon) \mathbb{S}_n  
+
e^{-ieVt} Y(t, \varepsilon + eV) \Gamma_{eq}(\varepsilon + eV/2) \hat P_+
\\+
e^{ieVt} Y(t, \varepsilon - eV) \Gamma_{eq} (\varepsilon - eV/2) \hat P_-
 =1.
\label{YVeq}
\end{multline}
The solution of this equation can be expressed in terms of the infinite series 
\begin{equation}
Y(t,\varepsilon)
=
\sum_{n=-\infty}^{\infty}
e^{ineVt} y_n(\varepsilon - neV),
\label{Ydef}
\end{equation}
where the coefficients $y_n(\varepsilon)$ obey the following recurrence relations 
\begin{multline}
y_n(\varepsilon) \mathbb{S}_n 
+
y_{n+1} (\varepsilon) \Gamma_{eq}(\varepsilon + neV + eV/2) \hat P_+
\\+
y_{n-1} (\varepsilon) \Gamma_{eq}(\varepsilon + neV - eV/2) \hat P_- = \delta_{n,0}.
\label{yn}
\end{multline}
Representing the coefficients $y_n(\varepsilon)$ in terms of the products
\begin{gather}
y_n(\varepsilon)
=
\begin{cases}
\alpha_0(\varepsilon)\alpha_1(\varepsilon) \cdots \alpha_n(\varepsilon),& n \geqslant 0, 
\\
\alpha_0(\varepsilon)\alpha_{-1}(\varepsilon) \cdots \alpha_n(\varepsilon),& n \leqslant 0,
\end{cases}
\end{gather}
one arrives at the recurrence relations for the coefficients $\alpha_n$ which read
\begin{multline}
\alpha_n(\varepsilon) 
=
-
\Gamma_{eq}(\varepsilon + neV - eV/2) \hat P_-
\\\times
[\mathbb{S}_n + \alpha_{n+1}(\varepsilon) \Gamma_{eq}(\varepsilon + neV + eV/2) \hat P_+]^{-1},
\quad n >0,
\label{ap}
\end{multline}
\begin{multline}
\alpha_{n}(\varepsilon)
=
-\Gamma_{eq}(\varepsilon + neV + eV/2) \hat P_+
\\\times
[\mathbb{S}_n
+
\alpha_{n-1}(\varepsilon) \Gamma_{eq}(\varepsilon + neV - eV/2) \hat P_-]^{-1},
\quad n < 0
\label{an}
\end{multline}
and
\begin{multline}
\alpha_0(\varepsilon)
=
[\mathbb{S}_n 
+
\alpha_{-1}(\varepsilon) \Gamma_{eq}(\varepsilon - eV/2) \hat P_-
\\+
\alpha_{1}(\varepsilon) \Gamma_{eq}(\varepsilon + eV/2) \hat P_+]^{-1}.
\label{a0}
\end{multline}
Here the coefficients $\alpha_{n}(\varepsilon)$ are off-diagonal matrices in the Nambu space making $y_n$ diagonal (off-diagonal) matrices for even (odd) indices $n$. We also note that the coefficients $\alpha_{n}(\varepsilon)$ tend to zero for large $|n|$. This property can be conveniently employed, e.g., in numerical calculations.

Once the coefficients $y_n$ are found, one can immediately evaluate ${\mathcal T}$ and the Josephson current. We get 
\begin{equation}
I (t)=\sum_{n=-\infty}^{\infty} I_n (V) e^{i n \varphi(t)},
\label{CPRNE}
\end{equation}
where $\varphi (t)$ is defined in Eq. \eqref{phiL} and 
\begin{multline}
I_n (V) =  
\frac{e}{8}\sum_{n'=-\infty}^{\infty}
\int \dfrac{d \varepsilon}{2\pi}
\Sp
\Bigl[
y_{2n+n'}(\varepsilon - (2n+n') eV)
\\\times
Q(\varepsilon, eV)
y_{n'}^+(\varepsilon - n' eV)
\left(
\mathbb{S}_n^+
\hat o_3 \hat \tau_3
\mathbb{S}_n 
-
\hat o_3 \hat \tau_3
\right) 
\Bigr].
\label{gKy}
\end{multline}

This is the key result of our present paper. We observe that the current $I$ defined in Eqs. \eqref{CPRNE}, \eqref{gKy} is manifestly $2\pi$-periodic in $\varphi$ thus leaving no room for any  speculations about the presence of "$4\pi$-periodic ac Josephson effect" in superconducting junctions. We would also like to emphasize that no approximations have been performed while deriving Eqs. \eqref{CPRNE}, \eqref{gKy}, i.e. these equations are exact for the model considered here. They effectively generalize $2\pi$-periodic CPR in Eq. \eqref{3} to non-stationary and non-equilibrium situations provided the superconducting phase $\varphi$ depends linearly on time.

\section{Spin-active scatterer}

Equations \eqref{CPRNE}, \eqref{gKy} describing ac Josephson effect are fairly general embracing a wide range of superconducting weak links. Below we will consider an important example of a superconducting junction with magnetic scattering. In order to specify the corresponding model we will assume that quasiparticles with opposite spin orientations scatter independently as they propagate between superconductors. Then the scattering matrix takes the form
\begin{equation}
[\mathbb{S}_n]_{\sigma}
=
\begin{pmatrix}
\sqrt{R_{\sigma}}  & -i \sqrt{D_{\sigma}}  &  0 & 0
\\
-i \sqrt{D_{\sigma}}  & \sqrt{R_{\sigma}}  & 0 & 0 \\
0 & 0 & \sqrt{R_{-\sigma}}  & i \sqrt{D_{-\sigma}}
\\
0 & 0 & i \sqrt{D_{-\sigma}}  & \sqrt{R_{-\sigma}}  \\
\end{pmatrix}
e^{-i\theta_{\sigma}/2}.
\label{Ssfsfun}
\end{equation}
Here $\sigma$ accounts for the spin variable, $\theta_{\sigma} = \theta \hat \sigma_3$ represents the so-called spin-mixing angle, $D_+$ and $D_-$ define the transmission values respectively for spin-up and spin-down quasiparticles, $R_{\pm} = 1 - D_{\pm}$ are the corresponding reflection coefficients and  $\hat \sigma_3$ is the Pauli matrix. For simplicity, below we will also assume that our superconducting junction is symmetric and set $|\Delta_1|=|\Delta_2|=|\Delta|$.

\subsection{Several limiting cases}

We first briefly address the limit of normal junctions by setting $|\Delta| \to 0$. In this case all the coefficients $y_n$ in Eq. \eqref{gKy} vanish except for $y_0$ which is equal to $\mathbb{S}_n^+$. As a result, for ${\mathcal T}$  \eqref{gKy} we obtain
\begin{equation}
{\mathcal T}  = 
\dfrac{eV}{2\pi}
\Sp
\left(
1
-
\hat o_3 \hat \tau_3
\mathbb{S}_n
\hat o_3 \hat \tau_3
\mathbb{S}_n^+
\right),
\label{gKynorm}
\end{equation}
Combining Eqs. \eqref{Ssfsfun}, \eqref{gKynorm} and  \eqref{onech}, we immediately recover the standard result
\begin{equation}
I=G_NV, \quad G_N  = 
\dfrac{e^2}{h}(D_+ + D_-),
\end{equation}
where $h=2\pi$ in our notations.

Turning on superconductivity and employing Eq. \eqref{Ssfsfun} we recostruct the spectrum of the corresponding subgap Andreev states with energies
\begin{equation}
\varepsilon_{\mu_1, \mu_2}(\varphi) = |\Delta| \mu_1 \cos[(\theta - \mu_2 \eta)/2].
\label{Andst}
\end{equation}
where the quantum numbers $\mu_{1,2}=\pm 1$ and the parameter $\eta$ ($0< \eta<\pi$) is unambiguously determined from the following equation
\begin{equation}
\cos \eta = \sqrt{R_+ R_-} + \sqrt{D_+ D_-} \cos \varphi.
\label{coseta}
\end{equation}
In equilibrium, the Josephson phase $\varphi$ does not depend on time and the current can be conveniently evaluated making use of Eq. \eqref{IS}. Combining this equation with Eqs. \eqref{Andst}, \eqref{coseta} we arrive at the well known equilibrium CPR \cite{Barash} derived for magnetic junctions.

In the case of a small but non-zero bias voltage $V \ll |\Delta |/e$ applied to the junction our exact result in Eqs. \eqref{CPRNE}, \eqref{gKy} allows to dismiss the claim \cite{Chalmers} of the existence of the "$4\pi$-periodicity of the ac Josephson current" under the condition $\sin (\theta /2) >\sqrt{D}$ (assuming that $D_+=D_-=D$). Note that the latter condition indeed assures that $4\pi$-periodic in $\varphi$ Andreev levels \eqref{Andst}, \eqref{coseta} never touch the continuum which, however, does not yet constitute the case for $4\pi$-periodicity of the Josephson current -- contrary to the scenario outlined in the Introduction. We will return to this issue towards the end of the paper. 

Here we only point out that Eq. \eqref{dEdf} employed within that scenario holds only in equilibrium and is in general unsuitable for calculation of the current provided a non-zero bias voltage is applied to the junction. For non-magnetic junctions with non-zero reflection coefficient this statement is illustrated, e.g., by the results \cite{GZ23}. Below we will consider a special limit of fully transparent junctions which is also of interest in that respect.

\subsection{Fully transparent junctions}
\begin{figure}
\includegraphics[width=80mm]{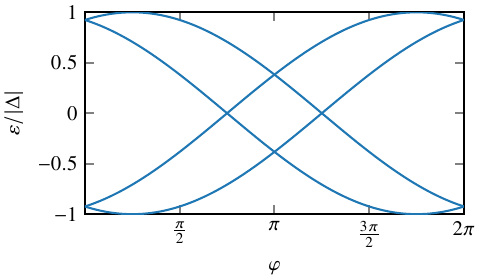}
\caption{The phase dependent energies of four subgap  Andrev bound states in transparent magnetic junctions. The spin-mixing angle $\theta$ is set equal to $\theta=\pi/4$.}
\label{abs-fig}
\end{figure}

Let us set $D_{\pm} = 1$ and, as before, $|\Delta_{1,2}|=|\Delta|$.  Then the normal state scattering matrix $\mathbb{S}_n$ \eqref{Ssfsfun} obeys the conditions
\begin{equation}
\hat P_+ \mathbb{S}_n \hat P_+ =\hat P_- \mathbb{S}_n \hat P_-=0
\label{Strans}
\end{equation}
and our recurrence relations get drastically simplified. We obtain
\begin{gather}
\alpha_n(\varepsilon) 
=
-
\Gamma_{eq}(\varepsilon + neV - eV/2) \hat P_-
\mathbb{S}_n^+,
\quad n >0,
\label{aptrans}
\\
\alpha_{n}(\varepsilon)
=
-\Gamma_{eq}(\varepsilon + neV + eV/2) \hat P_+
\mathbb{S}_n^+,
\quad n < 0,
\label{antrans}
\\
\alpha_{0}(\varepsilon)=\mathbb{S}_n^+.
\end{gather}
As a result, for $y_n(\varepsilon)$ we find 
\begin{gather}
\begin{split}
y_n(\varepsilon - neV)
= &
\mathbb{S}_n^+
(-\hat \tau_1 \hat P_- \mathbb{S}_n^+)^n
\\\times &
\prod_{k=1}^n
a(\varepsilon - keV + eV/2),
\quad n>0,
\end{split}
\\
\begin{split}
y_n(\varepsilon - neV)
= &
\mathbb{S}_n^+
(-\hat \tau_1 \hat P_+ \mathbb{S}_n^+)^{|n|}
\\\times &
\prod_{k=-1}^n
a(\varepsilon - keV - eV/2),
\quad n<0,
\end{split}
\\
y_0(\varepsilon)=\mathbb{S}_n^+.
\end{gather}
Here scattering matrix is parameterized by a single parameter $\theta$  
\begin{equation}
\mathbb{S}_n = -i \hat o_1 \hat \tau_3 e^{-i\theta \hat \sigma_3 /2}.
\label{Stransdef}
\end{equation}
Then for products $(-\hat \tau_1 \hat P_{\pm} \mathbb{S}_n^+)^n$ we get 
\begin{equation}
(-\hat \tau_1 \hat P_{\pm} \mathbb{S}_n^+)^n
=
P_{\mp} (\hat o_2 \tau_2)^n e^{i n\theta \hat \sigma_3 /2}, \quad n>0.
\end{equation}
Making use of the above equations together with Eq. \eqref{Ydef} one can evaluate the function  $Y(t, \varepsilon)$. Here, however, we will proceed slightly differently and evaluate this function directly from Eq. \eqref{YVeq}. Employing \eqref{Stransdef}, we find 
\begin{multline}
Y (t, \varepsilon) \hat o_1 \hat \tau_2 e^{-i\theta \hat \sigma_3 /2}
+
e^{-ieVt} Y(t, \varepsilon + eV) a(\varepsilon + eV/2) \hat P_-
\\+
e^{ieVt} Y(t, \varepsilon - eV) a(\varepsilon - eV/2)  \hat P_+
 = \hat \tau_1,
\label{YVeqtrans}
\end{multline}
The matrix $Y (t, \varepsilon)$ can be represented as a sum of eight terms, all having a different matrix structure:
\begin{multline}
Y (t, \varepsilon) 
=
\hat \tau_1 \sum_{s_1,s_2,s_3 = \pm} Y_{s_1, s_2, s_3} (t, \varepsilon) 
\\\times
\dfrac{1 + s_1 \hat\sigma_3}{2}
\dfrac{1 + s_2 \hat o_1 \hat \tau_2}{2}
\dfrac{1 + s_3 \hat o_3 \hat \tau_3}{2}.
\end{multline}
With the aid of this representation one can rewrite Eq. \eqref{TVconst} for the function ${\mathcal T}$ in terms of the functions $Y_{s_1, s_2, s_3}$. We obtain
\begin{multline}
{\mathcal T} =
2 \sum_{s_1,s_2,s_3 = \pm}
s_3
\int 
\dfrac{d \varepsilon}{2\pi}
|Y_{s_1, s_2, s_3} (t, \varepsilon)|^2
\\\times
[1 - |a(\varepsilon + s_3 eV/2)|^2 ] h_0(\varepsilon + s_3 eV/2),
\label{TVconsttrans}
\end{multline}
where 
\begin{equation}
a(\varepsilon)= \dfrac{-\varepsilon + \sqrt{\varepsilon^2 - |\Delta|^2}}{|\Delta|},
\end{equation}
is the Andreev amplitude. Making use the fact that function $a(\varepsilon)$ varies slowly on the energy scale $eV$  (see Appendix \ref{AY}) we obtain with a good accuracy
\begin{equation}
Y_{s_1, s_2, s_3}  (t, \varepsilon)
=
\sum_{k=0}^{\infty}
e^{\Phi_{s_1, s_2, s_3} (t, \varepsilon, V, k)},
\label{Ys}
\end{equation} 
where
\begin{multline}
\Phi_{s_1, s_2,s_3} (t, \varepsilon, V, k)
=
\dfrac{s_3}{eV}
\int_{\varepsilon-ks_3eV}^{\varepsilon}
\ln[ a(\varepsilon_1)
] d \varepsilon_1 
\\+
i ks_3eV t
+
i \dfrac{\theta s_1 (k+1) }{2}
+
i\pi (k+1) \dfrac{1+s_2}{2}.
\label{phis}
\end{multline}
The real part of the function $\Phi_{s_1, s_2,s_3}$ vanishes provided the energy tends to infinity 
\begin{equation}
\Phi_{s_1, s_2,s_3} (t, \varepsilon, V, k) \rightarrow 0,
\quad |\varepsilon| \rightarrow \infty.
\end{equation}
The phases of the terms in Eq. \eqref{Ys} strongly depend on the summation index $k$, thus resulting in their cancellation. The main contribution to the sum in Eq. \eqref{Ys} is provided by the index values for which the function $e^{i\Img \Phi}$ depends weakly on $k$, i.e. in the vicinity of the points determined by the condition 
\begin{multline}
\dfrac{\partial \Img \Phi_{s_1, s_2,s_3} (t, \varepsilon, V, k) }{ \partial k}
=
\Img\ln[ a(\varepsilon-ks_3eV)]
\\+
s_3eV t
+
\dfrac{\theta s_1 }{2}
+
\pi \dfrac{1+s_2}{2}
=
2\pi N,
\label{PhiK}
\end{multline}
where $N$ is an arbitrary integer number. Under the condition $|\varepsilon-ks_3eV|<|\Delta|$ Eq. \eqref{PhiK} just defines the energies of Andreev bound states
\begin{equation}
\varepsilon_{s_1,s_3}=
|\Delta|
\cos(s_3eV t + \theta s_1 /2) \sgn [\sin (s_3eV t + \theta s_1 /2)].
\label{abs}
\end{equation}
The energy dependence on the phase $\varphi =2eVt$ for four Andreev bound states \eqref{abs} is displayed in Fig. \ref{abs-fig}. Note that in the limit $\theta \to 0$ the states for different spin orientations merge pairwise forming two (instead of four) Andreev levels which coincide with those for transparent non-magnetic junctions.

Note that Eq. \eqref{PhiK} has a solution only for particular values of the parameters $s_1$, $s_2$, $s_3$. From the condition $0 \leqslant\Img \ln a(\varepsilon) \leqslant \pi$ we observe that for given values $s_1$ and $s_3$ Eq. \eqref{PhiK} has exactly one solution that fixes both values of $k$ and the parameter $s_2$. The latter then reads 
\begin{equation}
s_2 =  \sgn \sin (s_3eV t + \theta s_1 /2).
\label{s2}
\end{equation}

Evaluating $|Y_{s_1, s_2, s_3}  (t, \varepsilon)|^2$ (see Appendix \ref{AY}) and combining Eqs. \eqref{TVconsttrans} and \eqref{IS} with the resulting expression \eqref{Ysss4} derived under the condition
\begin{equation}
1 - |\varepsilon_{s_1,s_3}/\Delta | \gg |eV/\Delta |^{2/3},
\label{cond1}
\end{equation}
we arrive at the final result in the form
\begin{multline}
I (t) = 
\frac{I_c}{2}
\Bigl[|\sin(eVt + \theta/2)| + |\sin(eVt - \theta/2)| \Bigr]\sgn V ,
\label{TVconsttrans2}
\end{multline}
where 
\begin{equation}
I_c=e|\Delta| \tanh\frac{|\Delta|}{2T}
\end{equation}
is the critical current for a transparent single channel superconducting weak link.

For $\theta \to 0$ the result \eqref{TVconsttrans2} reduces to that of Averin and Bardas \cite{AB1,AB2}. Having derived Eq. \eqref{TVconsttrans2} from our rigorous calculation we observe that the same expression for the current would follow if we simply sum up the derivatives of the energies \eqref{abs} with respect to the phase $\varphi =2eVt$ (cf. Eq. \eqref{dEdf}) for both spin directions within the intervals $0<eVt \pm \theta/2 <\pi$ (and periodically extended otherwise). It follows from Eq. \eqref{cond1}, however, that the actual validity domain of the result  \eqref{TVconsttrans2} shrinks considerably being restricted to the energy values sufficiently far from the points $\varepsilon_{s_1,s_3}=\pm |\Delta|$ where Andreev bound states touch the continuum. 

In the vicinity of these points, i.e. provided the condition Eq. \eqref{cond1} is violated and the arguments of the $\sin$ terms in Eq. \eqref{TVconsttrans2} get sufficiently close to $\pi n$, one needs to set up an extra calculation which is presented in Appendix \ref{AY}. It yields
\begin{multline}
I (t) = 
\frac{I_c}{2}\left(\dfrac{|eV|}{|\Delta|}\right)^{1/3}\sgn(V) 
\\\times
\sum_\pm F\left( 2|\sin (eVt \pm \theta/2)| \left[\dfrac{|\Delta|}{|eV|}\right]^{1/3}\right),
\label{TVconsttrans3}
\end{multline}
where $F(y)$ is a universal function of order unity at $y \lesssim 1$ and $F(y) \simeq y/2$ for $y \gg 1$ implying that 
Eq. \eqref{TVconsttrans3} reduces back to Eq. \eqref{TVconsttrans2} in the limit \eqref{cond1}, i.e. as soon as Andreev levels move far from the gap edges. On the other hand, at every period of Josephson oscillations under the condition $-|eV/\Delta |^{1/3} \lesssim eVt \pm \theta /2 -\pi n \lesssim |eV/\Delta |^{1/3}$ the current $I$ strongly deviates from that in Eq. \eqref{TVconsttrans2} acquiring a dissipative component $\propto  |V|^{1/3}$. Averaging this periodic dissipative current over time we recover a sub-Ohmic contribution to the $I-V$ curve $\propto |V|^{2/3}$.

Note that very recently a similar result was derived for non-magnetic transparent weak links within a different technique \cite{GZPisma,GZ24}. In the limit $|eV| \ll |\Delta |$ the corresponding $I-V$ curve reads \cite{GZPisma,GZ24}
\begin{equation}
\bar{I}=\frac{2}{\pi}I_c\Bigl[1+ 0.59 \left(\dfrac{|eV|}{|\Delta|}\right)^{2/3}\Bigr]\sgn V ,
\label{IVcurve}
\end{equation} 
where the first term represents the excess current \cite{Gunsenheimer94} while the second one accounts for a sub-Ohmic dissipative contribution to $\bar{I}$. Averaging Eq. \eqref{TVconsttrans3} over time we observe that the resulting $I-V$ curve does not depend on the spin-mixing angle $\theta$ and, hence, should coincide \cite{FN2} with Eq. \eqref{IVcurve}.

Finally, let us not that 'periodic dissipation' discussed here is somewhat reminiscent of the well known '$\cos\varphi$' dissipative contribution to the current across Josephson tunnel barriers \cite{BP}. An important difference between the two, however, is that the '$\cos\varphi$'-term only appears at higher voltages $eV >2|\Delta |$, whereas here -- due to the effect of MAR -- we are dealing with non-vanishing dissipation already at small voltages $eV \ll |\Delta |$.

\section{Discussion: Non-unitary evolution of Andreev states}
For symmetric junctions with $|\Delta_1| = |\Delta_2| = |\Delta|$ our general expression for the current defined in  Eqs. \eqref{onech}, \eqref{T3} can also be reformulated in a somewhat different manner. For this purpose it is convenient to introduce the matrix
\begin{equation}
W(t) = \tilde{\mathbb{S}}_n (t) \hat \tau_1,
\label{Wt}
\end{equation}
where $\hat \tau_1$ is the corresponding Pauli matrix and the scattering matrix  $\tilde{\mathbb{S}}_n (t)$ is defined in Eq. \eqref{Ssfsfuntild} of Appendix \ref{AW} for an arbitrary time dependence of the Josephson phase $\varphi (t)$. The matrix \eqref{Wt} obeys the symmetry relation 
\begin{equation}
W = \hat o_1 \hat \sigma_2 W^+ \hat \sigma_2 \hat o_1 = \hat o_1 \hat \sigma_1 W^+ \hat \sigma_1 \hat o_1
\label{Wsym}
\end{equation}
following directly from Eq. \eqref{Ssym}. After performing a unitary transformation 
\begin{equation}
\tilde W = U W U^+,
\end{equation}
with the time independent unitary matrix 
\begin{equation}
U = 
\dfrac{1}{2\sqrt{2}}
[(1 + \hat \sigma_1)(1 + \hat \sigma_3) 
-
(1 - \hat \sigma_1)(1 - \hat \sigma_3) \hat o_1 ]
\end{equation}
the matrix $\tilde W$ splits into blocks
\begin{equation}
\tilde W
=
\begin{pmatrix}
W_{++} & W_{+-} \\
W_{-+} & W_{--}
\end{pmatrix}.
\label{blocks}
\end{equation}

Making use of the relation
\begin{equation}
\Gamma_{eq}(\varepsilon) = \hat \tau_1 a(\varepsilon),
\end{equation}
we rewrite the general expression for the current defined by Eqs. \eqref{T3} and \eqref{onech} in the following equivalent form
\begin{multline}
I =
\frac{ie}{2}
\Sp
\Bigl[
(\tilde W + a)^{-1} 
(H_{eq} - a H_{eq} a^+)
\\\times
(\tilde W^+ + a^+ )^{-1}
\tilde W^+
\partial_{\varphi}\tilde W
\Bigr],
\label{TW}
\end{multline}
where $a$ is a retarded nonlocal integral operator with the kernel 
\begin{equation}
a(t-t') = 
i \dfrac{J_1[|\Delta|(t-t')]}{t-t'} \Theta(t-t')
\end{equation}
and $J_1$ is Bessel function of the first kind.

Consider, e.g., the inverse operator $(\tilde W + a)^{-1}$ in Eq. \eqref{TW}. It can be identically rewritten as 
\begin{multline}
(\tilde W +  a)^{-1}
=
-\dfrac{|\Delta|}{2}
\begin{pmatrix}
(\varepsilon - \mathcal{H}_{+})^{-1} & 0 \\
0 & (\varepsilon - \mathcal{H}_{-})^{-1}
\end{pmatrix}
\\\times
\Biggl[
\begin{pmatrix}
W_{+-} a^{-1} W_{+-}^{-1} & 0 
\\
0 & W_{-+} a^{-1} W_{-+}^{-1}
\end{pmatrix} \tilde W^+ + 1
\Biggr] ,
\label{inv2}
\end{multline}
where we introduced two effective "Hamiltonians" $\mathcal{H}_\pm$. One can verify that in the case of the scattering matrix 
in the form \eqref{Ssfsfun} considered here the matrices in Eq. \eqref{blocks} obey additional symmetry relations
\begin{equation}
W_{++} = W_{--}, \quad  W_{+-} = W_{-+}.
\label{sympm}
\end{equation}
Under the condition \eqref{sympm} the two "Hamiltonians" coincide with each other $\mathcal{H}_+=\mathcal{H}_-=\mathcal{H}$ and read
\begin{multline}
\mathcal{H} = \dfrac{|\Delta|}{2}[
W_{++}
+
W_{+-} a^{-1} W_{+-}^{-1} W_{++} a
\\+
W_{+-} a^{-1} W_{+-}^{-1} - a^{-1}
] ,
\label{Ham}
\end{multline} 
where the matrices $W_{++}$ and $W_{+-}$ are explicitly defined in Eqs. \eqref{B8}-\eqref{B15} of Appendix \ref{AW}. It is also worth emphasizing that no approximation was performed while deriving the above equations, i.e. the representation of the inverse operator \eqref{inv2}-\eqref{Ham} is exact for the model considered here. 

Let us analyze the expression \eqref{Ham}.  In equilibrium, i.e. for $V \equiv 0$ and the time independent Josephson phase $\varphi$, the matrices $W_{++}$ and $W_{+-}$ are also time independent and, hence, they both commute with the Andreev amplitude operator $a$ which then drops out from the right-hand side of Eq. \eqref{Ham}. It follows immediately that in equilibrium the Hamiltonian $\mathcal{H}$ reduces to the Hermitian matrix $\mathcal{H}=|\Delta|W_{++}$ which spectrum coincides with that of Andreev subgap bound states for our problem. Accordingly, in this particular limit the standard quantum mechanical treatment of Andreev states is justified and appropriate. 

The situation changes drastically as soon as non-zero external voltage bias $V$ is turned on. No matter how small $V$ is, the Josephson phase $\varphi$ as well as the matrices $W_{++}$ and $W_{+-}$ now explicitly depend on time and, hence, $\mathcal{H}$ becomes a non-local in time retarded integral operator describing {\it non-unitary} evolution of Andreev states. Non-locality in time generally implies both dissipation \cite{SZ90} and dephasing \cite{GZ98,GZS}, thus making the standard quantum mechanical analysis insufficient. Hence, manipulating with Andreev states just like with ordinary quantum mechanical levels may yield unreliable results even in the so-called "adiabatic" limit $0<eV \ll |\Delta|$, as it is actually demonstrated by the results derived here as well as in Ref. \onlinecite{GZ23}.

Non-unitary evolution of Andreev states implies that at any non-zero $V$ such states become unstable due to the presence of electric field inside a weak link and the effect of MAR. As a result, even at small $V$ electrons and holes may -- depending on their velocity 
directions -- significantly increase or decrease their energies while moving in-between two superconductors, thus making Andreev level quantization not anymore possible. For this reason it is dangerous to rely on an oversimplified physical picture of Andreev states at equilibrium while describing ac Josephson effect in superconducting weak links. The latter description generally requires a complete microscopic many-body calculation properly taking into account all non-equilibrium effects. This kind of a calculation was carried out in our present work.

\vspace{0.5cm}

\centerline{\bf Acknowledgement}

\vspace{0.5cm}
We are grateful to A.V. Galaktionov for illuminating discussions.

\appendix 

\section{Matrix relations between Riccati amplitudes}
\label{Aric}
Matrix relations between Riccati amplitudes \eqref{Gamma1} and \eqref{Gamma2} can also be represented in a number of different though equivalent ways. We have
\begin{gather}
(1 - \hat{\tilde \Gamma}^R \hat \gamma^R)^{-1}
(\hat{\tilde \Gamma}^R  - \hat{\tilde a}^R )
=
(\underline{\mathcal{S}} 
-
\hat{\tilde \gamma}^R
\mathcal{S}
\hat \gamma^R)^{-1}
\hat{\tilde \gamma}^R
\mathcal{S},
\label{Gamma3}
\\
(1 - \hat{\tilde \gamma}^A \hat a^A)(1 - \hat{\tilde \gamma}^A \hat \Gamma^A)^{-1}
=
\underline{\mathcal{S}}^+
(\underline{\mathcal{S}}^+
-
\hat{\tilde \gamma}^A
\mathcal{S}^+
\hat \gamma^A
 )^{-1},
\label{Gamma4}
\\
(1 - \hat \gamma^A \hat{\tilde a}^A)
(1 - \hat \gamma^A \hat{\tilde \Gamma}^A)^{-1}
=
\mathcal{S}
(\mathcal{S}
-
\hat \gamma^A
\underline{\mathcal{S}}
\hat{\tilde \gamma}^A
)^{-1},
\label{Gamma5}
\\
(\hat \Gamma^A - \hat a^A)(1 - \hat{\tilde \gamma}^A \hat \Gamma^A)^{-1}
=
\mathcal{S}^+
\hat \gamma^A
(\underline{\mathcal{S}}^+
-
\hat{\tilde \gamma}^A
\mathcal{S}^+
\hat \gamma^A )^{-1},
\label{Gamma6}
\\
(1 - \hat{\tilde \gamma}^R \hat \Gamma^R)^{-1}
(1 - \hat{\tilde \gamma}^R \hat a^R)
=
\underline{\mathcal{S}}
(
\underline{\mathcal{S}}
-
\hat{\tilde \gamma}^R
\mathcal{S}
\hat \gamma^R )^{-1} ,
\label{Gamma7}
\\
(1 - \hat \gamma^R \hat{\tilde \Gamma}^R)^{-1}
(1 - \hat \gamma^R \hat{\tilde a}^R)
=
\mathcal{S}^+
(
\mathcal{S}^+
-
\hat \gamma^R
\underline{\mathcal{S}}^+
\hat{\tilde \gamma}^R )^{-1} ,
\label{Gamma8}
\\
(1 - \hat \Gamma^R \hat{\tilde \gamma}^R)^{-1}
(\hat \Gamma^R-\hat a^R)
=
(
\mathcal{S}^+
-
\hat \gamma^R
\underline{\mathcal{S}}^+
\hat{\tilde \gamma}^R
)^{-1}
\hat \gamma^R
\underline{\mathcal{S}}^+,
\label{Gamma9}
\\
(1 - \hat a^A \hat{\tilde \gamma}^A)(1 - \hat \Gamma^A \hat{\tilde \gamma}^A)^{-1}
=
(\mathcal{S}
-
\hat \gamma^A
\underline{\mathcal{S}}
\hat{\tilde \gamma}^A )^{-1}
\mathcal{S},
\label{Gamma10}
\\
(1 - \hat{\tilde a}^A \hat \gamma^A)
(1 - \hat{\tilde \Gamma}^A \hat \gamma^A)^{-1}
=
(\underline{\mathcal{S}}^+
-
\hat{\tilde \gamma}^A
\mathcal{S}^+
\hat \gamma^A )^{-1}
\underline{\mathcal{S}}^+,
\label{Gamma11}
\\
(\hat{\tilde \Gamma}^A - \hat{\tilde a}^A)
(1 - \hat \gamma^A \hat{\tilde \Gamma}^A)^{-1}
=
\underline{\mathcal{S}}
\hat{\tilde \gamma}^A
(\mathcal{S}
-
\hat \gamma^A
\underline{\mathcal{S}}
\hat{\tilde \gamma}^A  )^{-1}.
\label{Gamma12}
\end{gather}
The above equations may be useful for practical calculations.

\section{Block matrix inversion}
The square matrix formed by four square sub-matrices can be inverted blockwise as
\begin{multline}
\begin{pmatrix}
A & B \\
C & D
\end{pmatrix}^{-1}
=
\begin{pmatrix}
(A - B D^{-1} C)^{-1} & (C - D B^{-1} A)^{-1} \\
( B - A C^{-1} D)^{-1} & (D - C A^{-1} B)^{-1}
\end{pmatrix}
=\\=
\begin{pmatrix}
(A - B D^{-1} C)^{-1} & 0 \\
0 & (D - C A^{-1} B)^{-1}
\end{pmatrix}
\times\\\times
\begin{pmatrix}
1 & - B D^{-1} \\
- C A^{-1} & 1
\end{pmatrix}.
\label{blockinv}
\end{multline}

\section{Evaluation of the $Y$-function}
\label{AY}
The coefficients $Y_{s_1, s_2, s_3}$ obey the following relations 
\begin{multline}
Y_{s_1, s_2, s_3} (t, \varepsilon) 
s_2 e^{-i\theta s_1 /2}
\\+
e^{is_3eVt} 
Y_{s_1, s_2, s_3} (t, \varepsilon - s_3 eV) 
a(\varepsilon - s_3 eV/2) 
=1.
\label{YVeqtrans2}
\end{multline}

Let us first construct a general solution for the corresponding homogeneous equation. This solution can be expressed in the form 
\begin{equation}
Y_{s_1, s_2, s_3}^{hom}  (t, \varepsilon)
=
e^{P_{s_1, s_2, s_3}  (t, \varepsilon)/(s_3 eV)} p_{s_1, s_2, s_3}  (t, \varepsilon),
\end{equation}
where both $P$ and $p$ are some smooth functions of $\varepsilon$. We find 
\begin{multline}
Y_{s_1, s_2, s_3}  (t, \varepsilon)
=
C 
\exp\left(
i \varepsilon t
+
i \dfrac{\theta s_1 s_3\varepsilon }{2 eV}
+
i\pi s_3 \dfrac{1+s_2}{2eV}\varepsilon
\right)
\\\times
\exp\left(
\dfrac{s_3}{eV}
\int_{\varepsilon_0}^{\varepsilon}
\ln[ a(\varepsilon_1)
] d \varepsilon_1 \right),
\label{Yhom}
\end{multline}
where $C$ is some constant.

Let us choose the branch of the logarithm with $0\leqslant \Img \ln a \leqslant \pi$. We will seek a particular solution for the inhomogeneous equation in the form \eqref{Yhom} with the energy dependent prefactor $C(\varepsilon)$. In this way we arrive at the equation for $C(\varepsilon)$ 
\begin{multline}
C(\varepsilon)
-
C(\varepsilon - s_3 eV)
=s_2 e^{i\theta s_1 /2}
\\\times
\exp\left(-
i \varepsilon t
-
i \dfrac{\theta s_1 s_3\varepsilon }{2 eV}
-
i\pi s_3\dfrac{1+s_2}{2eV}\varepsilon
\right)
\\\times
\exp\left(-
\dfrac{s_3}{eV}
\int_{\varepsilon_0}^{\varepsilon}
\ln[ a(\varepsilon_1)
] d \varepsilon_1 \right)
\end{multline}
which allows to establish Eqs. \eqref{Ys}, \eqref{phis}.

Near the stationary points the summation over $k$ in Eq. \eqref{Ys} can -- with a good accuracy -- be replaced by integration. Then we get 
\begin{multline}
|Y_{s_1, s_2, s_3}  (t, \varepsilon)|^2
\\\approx
\Biggl|
\int dk
\exp\Biggl(
\dfrac{s_3}{eV}
\int_{\varepsilon-ks_3eV}^{\varepsilon}
\ln[ a(\varepsilon_1)
] d \varepsilon_1 
-
k \ln a(\varepsilon_{s_1,s_3})
\Biggr)
\Biggr|^2.
\label{Ysss}
\end{multline}
Here the index $s_2$ is not independent being fixed by Eq. \eqref{s2}. Equation \eqref{Ysss} can be rewritten identically as 
\begin{multline}
|Y_{s_1, s_2, s_3}  (t, \varepsilon)|^2
\approx
\dfrac{1}{|eV|^2}
\\\times
\Biggl|
\int d \varepsilon_2
\exp\Biggl(
\dfrac{s_3}{eV}
\int_{\varepsilon-\varepsilon_2}^{\varepsilon}
\ln\left[ a(\varepsilon_1)/a(\varepsilon_{s_1,s_3})\right] d \varepsilon_1 
\Biggr)
\Biggr|^2.
\label{Ysss2}
\end{multline}

This expression needs to be evaluated in the limit of low voltages $eV \ll |\Delta |$. We first assume that Andreev bound states energies $\varepsilon_{s_1,s_3}$ \eqref{abs} obey the condition \eqref{cond1}, i.e. remain sufficiently far from the gap edges. In this case the main contribution to the integral over $\varepsilon_2$ in Eq. \eqref{Ysss2} comes from an immediate vicinity of the point $\varepsilon_2 = \varepsilon - \varepsilon_{s_1,s_3}$. Expanding the exponent in Eq. \eqref{Ysss2}, we find 
\begin{multline}
|Y_{s_1, s_2, s_3}  (t, \varepsilon)|^2
\approx
\dfrac{1}{|eV|^2}
\\\times
\Biggl|
\int d \varepsilon_2
\exp\Biggl(
\dfrac{s_3}{eV}
\int_{\varepsilon_{s_1,s_3}}^{\varepsilon}
\ln\left[ a(\varepsilon_1)/a(\varepsilon_{s_1,s_3})\right] d \varepsilon_1 
\\-
\dfrac{s_3}{2eV} \dfrac{a'(\varepsilon_{s_1,s_3})}{a(\varepsilon_{s_1,s_3})}
(\varepsilon_2 - \varepsilon + \varepsilon_{s_1,s_3})^2
\Biggr)
\Biggr|^2.
\label{Ysss3}
\end{multline}
Evaluating the above integral over $\varepsilon_2$, we get 
\begin{multline}
|Y_{s_1, s_2, s_3}  (t, \varepsilon)|^2
\approx
\dfrac{2\pi}{|eV|} |\sin(s_3eV t + \theta s_1 /2)|
\\\times
\exp\Biggl(
\dfrac{2s_3}{eV}
\int_{\varepsilon_{s_1,s_3}}^{\varepsilon}
\ln |a(\varepsilon_1)|  d \varepsilon_1 
\Biggr)
\theta[s_3 (\varepsilon - \varepsilon_{s_1,s_3})/eV ].
\label{Ysss4}
\end{multline}
Combining this expression with Eqs. \eqref{TVconsttrans} and \eqref{IS} we recover the result \eqref{TVconsttrans2}.

Now let us assume that the the condition \eqref{cond1} is violated, i.e. the energy of at least one of the Andreev levels gets sufficiently close to the gap energy $\varepsilon = \pm |\Delta|$. In this case the function $a(\varepsilon)$ cannot anymore be expanded in Taylor series near these energies and our calculation needs to be modified accordingly. 

For definiteness we assume that one of the Andreev level $\varepsilon_A$ with quantum number $s_3 = \sgn(eV)$ is close to the energy $\pm |\Delta|$. Combining Eqs. \eqref{TVconsttrans} and \eqref{Ysss2} one can express ${\mathcal T}$ as a sum of all eight contributions 
\begin{equation}
{\mathcal T} = \sum_{p=1}^8{\mathcal T}_p,
\end{equation}
where 
\begin{widetext}
\begin{gather}
\label{T1}
{\mathcal T}_1=
\dfrac{2 \sgn (eV)}{|eV|^2}
\int_{-\infty}^{\infty}\dfrac{d\varepsilon}{2\pi}
[1 - |a(\varepsilon + |eV|/2)|^2 ] h_0(\varepsilon + |eV|/2)
\Biggl|
\int_0^{\infty} d \varepsilon_2 
\exp\left(
\dfrac{1}{|eV|}
\int_{\varepsilon-\varepsilon_2}^{\varepsilon}
\ln\left[ a(\varepsilon_1)/a(\varepsilon_A)\right] d \varepsilon_1 
\right)
\Biggr|^2,
\\
{\mathcal T}_2=
\dfrac{2 \sgn (eV)}{|eV|^2}
\int_{-\infty}^{\infty}\dfrac{d\varepsilon}{2\pi}
[1 - |a(\varepsilon + |eV|/2)|^2 ] h_0(\varepsilon + |eV|/2)
\Biggl|
\int_0^{\infty} d \varepsilon_2 
\exp\left(
\dfrac{1}{|eV|}
\int_{\varepsilon-\varepsilon_2}^{\varepsilon}
\ln\left[ -a(\varepsilon_1)/a(\varepsilon_A)\right] d \varepsilon_1 
\right)
\Biggr|^2, 
\\
{\mathcal T}_3=
\dfrac{2 \sgn (eV)}{|eV|^2}
\int_{-\infty}^{\infty}\dfrac{d\varepsilon}{2\pi}
[1 - |a(\varepsilon + |eV|/2)|^2 ] h_0(\varepsilon + |eV|/2)
\Biggl|
\int_0^{\infty} d \varepsilon_2 
\exp\left(
\dfrac{1}{|eV|}
\int_{\varepsilon-\varepsilon_2}^{\varepsilon}
\ln\left[ a(\varepsilon_1)/a(-\varepsilon_A)\right] d \varepsilon_1 
\right)
\Biggr|^2, 
\\
{\mathcal T}_4=
\dfrac{2 \sgn (eV)}{|eV|^2}
\int_{-\infty}^{\infty}\dfrac{d\varepsilon}{2\pi}
[1 - |a(\varepsilon + |eV|/2)|^2 ] h_0(\varepsilon + |eV|/2)
\Biggl|
\int_0^{\infty} d \varepsilon_2 
\exp\left(
\dfrac{1}{|eV|}
\int_{\varepsilon-\varepsilon_2}^{\varepsilon}
\ln\left[ -a(\varepsilon_1)/a(-\varepsilon_A)\right] d \varepsilon_1 
\right)
\Biggr|^2, 
\\
{\mathcal T}_5=
-\dfrac{2 \sgn (eV)}{|eV|^2}
\int_{-\infty}^{\infty}\dfrac{d\varepsilon}{2\pi}
[1 - |a(\varepsilon - |eV|/2)|^2 ] h_0(\varepsilon - |eV|/2)
\Biggl|
\int_0^{\infty} d \varepsilon_2 
\exp\left(-
\dfrac{1}{|eV|}
\int_{\varepsilon+\varepsilon_2}^{\varepsilon}
\ln\left[ a(\varepsilon_1)/a(\varepsilon_A)\right] d \varepsilon_1 
\right)
\Biggr|^2, 
\\
{\mathcal T}_6=
-\dfrac{2 \sgn (eV)}{|eV|^2}
\int_{-\infty}^{\infty}\dfrac{d\varepsilon}{2\pi}
[1 - |a(\varepsilon - |eV|/2)|^2 ] h_0(\varepsilon - |eV|/2)
\Biggl|
\int_0^{\infty} d \varepsilon_2 
\exp\left(-
\dfrac{1}{|eV|}
\int_{\varepsilon+\varepsilon_2}^{\varepsilon}
\ln\left[ -a(\varepsilon_1)/a(\varepsilon_A)\right] d \varepsilon_1 
\right)
\Biggr|^2, 
\\
{\mathcal T}_7=
-\dfrac{2 \sgn (eV)}{|eV|^2}
\int_{-\infty}^{\infty}\dfrac{d\varepsilon}{2\pi}
[1 - |a(\varepsilon - |eV|/2)|^2 ] h_0(\varepsilon - |eV|/2)
\Biggl|
\int_0^{\infty} d \varepsilon_2 
\exp\left(-
\dfrac{1}{|eV|}
\int_{\varepsilon+\varepsilon_2}^{\varepsilon}
\ln\left[ a(\varepsilon_1)/a(-\varepsilon_A)\right] d \varepsilon_1 
\right)
\Biggr|, 
\\
{\mathcal T}_8=
-\dfrac{2 \sgn (eV)}{|eV|^2}
\int_{-\infty}^{\infty}\dfrac{d\varepsilon}{2\pi}
[1 - |a(\varepsilon - |eV|/2)|^2 ] h_0(\varepsilon - |eV|/2)
\Biggl|
\int_0^{\infty} d \varepsilon_2 
\exp\left(-
\dfrac{1}{|eV|}
\int_{\varepsilon+\varepsilon_2}^{\varepsilon}
\ln\left[ -a(\varepsilon_1)/a(-\varepsilon_A)\right] d \varepsilon_1 
\right)
\Biggr|^2, 
\end{gather}
 
All these integrals are of the same type, hence it suffices to demonstrate how to handle only one of them, say, ${\mathcal T}_1$ \eqref{T1}. 
Consider, for instance, the contribution to ${\mathcal T}_1$ from energies in the vicinity of the point $\varepsilon=+|\Delta|$. For this contribution we approximately have
\begin{multline}
{\mathcal T}_1^+=
\dfrac{2 \sgn (eV)}{|eV|^2}
h_0(|\Delta|)
\int_{0}^{\infty}\dfrac{d\varepsilon}{2\pi}
\dfrac{2 \sqrt{2}}{\sqrt{|\Delta|}} \sqrt{\varepsilon}
\Biggl|
\int_0^{\varepsilon} d \varepsilon_2 
\exp\Biggl(
\dfrac{1}{|eV|}
\int_{\varepsilon-\varepsilon_2}^{\varepsilon}
\ln\left[ a(|\Delta| + \varepsilon_1)/a(\varepsilon_A)\right] d \varepsilon_1 
\Biggr)
\\+
\int_{0}^{\infty} d \varepsilon_2 
\exp\Biggl(
\dfrac{1}{|eV|}
\int_{-0}^{\varepsilon}
\ln\left[ a(|\Delta| + \varepsilon_1)/a(\varepsilon_A)\right] d \varepsilon_1 
+
\dfrac{1}{|eV|}
\int_{-\varepsilon_2}^{0}
\ln\left[ a(|\Delta| + \varepsilon_1)/a(\varepsilon_A)\right] d \varepsilon_1 
\Biggr)
\Biggr|^2
\\=
\dfrac{2 \sgn (eV)}{|eV|^2}
h_0(|\Delta|)
\int_{0}^{\infty}\dfrac{d\varepsilon}{2\pi}
\dfrac{2 \sqrt{2}}{\sqrt{|\Delta|}} \sqrt{\varepsilon}
\Biggl|
\int_0^{\varepsilon} d \varepsilon_2 
\exp\Biggl(
\dfrac{1}{|eV|}
\Biggl[
[i\pi - \ln a(\varepsilon_A)] \varepsilon_2  - \dfrac{2\sqrt{2}}{3\sqrt{|\Delta|}}
[\varepsilon^{3/2} - \varepsilon_2^{3/2}] \Biggr]
\Biggr)
\\+
\int_{0}^{\infty} d \varepsilon_2 
\exp\Biggl(
\dfrac{1}{|eV|}
\Biggl[
[i\pi - \ln a(\varepsilon_A)] \varepsilon  - \dfrac{2\sqrt{2}}{3\sqrt{|\Delta|}}
\varepsilon^{3/2} \Biggr]
+
\dfrac{1}{|eV|}
\Biggl[
[i\pi - \ln a(\varepsilon_A)] \varepsilon_2  - i \dfrac{2\sqrt{2}}{3\sqrt{|\Delta|}}
\varepsilon_2^{3/2} \Biggr]
\Biggr)
\Biggr|^2,
\label{t1plus}
\end{multline}
where we made use of the following expressions 
\begin{gather}
\int_{|\Delta|}^{\varepsilon} \ln a(\varepsilon_1) d \varepsilon_1
=
\begin{cases}
i\pi (\varepsilon - |\Delta|) - \dfrac{2\sqrt{2}}{3\sqrt{|\Delta|}}
(\varepsilon - |\Delta|)^{3/2},
&
\quad 
\varepsilon  > |\Delta| ,\  \varepsilon - |\Delta| \ll |\Delta|
\\
i\pi (\varepsilon - |\Delta|) 
+i \dfrac{2\sqrt{2}}{3\sqrt{|\Delta|}}
(|\Delta| - \varepsilon)^{3/2} ,
&
\quad 
\varepsilon < |\Delta| ,\  |\Delta| - \varepsilon \ll |\Delta|,
\end{cases}
\\
1 - |a(\varepsilon)|^2 \approx 
\dfrac{2 \sqrt{2}}{\sqrt{|\Delta|}} \sqrt{|\varepsilon| - |\Delta|}\theta(|\varepsilon| - |\Delta|),
\quad 
|\varepsilon| - |\Delta| \ll |\Delta|,
\end{gather}
valid near the gap energies. 

\end{widetext}

Provided the Andreev state energy $\varepsilon_A$ approaches the gap edge $|\Delta|$  
one may set $\varepsilon_A = |\Delta| [1 - (\varphi - \varphi_0)^2/8]$. Then one finds
\begin{equation}
i\pi - \ln a(\varepsilon_{s_1,s_3}) \approx 
i \dfrac{\sqrt{2}}{\sqrt{|\Delta|}} \sqrt{|\Delta| - \varepsilon_A} = i
\dfrac{|\varphi - \varphi_0|}{2},
\end{equation}
It is convenient to introduce the new variables
\begin{equation}
\tilde\varepsilon = \varepsilon \left(\dfrac{|eV|}{|\Delta|}\right)^{-2/3},
\quad
\tilde\varepsilon_2 = \varepsilon_2 \left(\dfrac{|eV|}{|\Delta|}\right)^{-2/3},
\end{equation}
and make the integral in Eq. \eqref{t1plus} dimensionless. Then the contribution ${\mathcal T}_1^+$ takes the form 
\begin{multline}
{\mathcal T}_1^+ = \sgn(eV) |\Delta| h_0(|\Delta|)
\left(\dfrac{|eV|}{|\Delta|}\right)^{1/3}
\\\times
F_1^+\left[ |\varphi - \varphi_0| \left(\dfrac{|\Delta|}{|eV|}\right)^{1/3}\right],
\label{f1plus}
\end{multline}
where $F_1^+$ is some universal function. Using similar approach we can evaluate all other contributions ${\mathcal T}_i^{\pm}$. They have exactly the same behavior as ${\mathcal T}_1^+$ with different but similarly behaving universal functions $F_i^{\pm}$. Collecting all the contributions we arrive at Eq. \eqref{TVconsttrans3}.

\section{$W$-matrices}
\label{AW}

For the model considered here the scattering matrix  $\tilde{\mathbb{S}}_n (t)$ takes the form 
\begin{widetext}
\begin{equation}
[\tilde{\mathbb{S}}_n]_{\sigma}
=
\begin{pmatrix}
\sqrt{R_{\sigma}}  & -i \sqrt{D_{\sigma}} e^{-i\varphi (t)/2}  &  0 & 0
\\
-i \sqrt{D_{\sigma}} e^{i\varphi (t)/2}  & \sqrt{R_{\sigma}}  & 0 & 0 \\
0 & 0 & \sqrt{R_{-\sigma}}  & i \sqrt{D_{-\sigma}} e^{i\varphi (t)/2}
\\
0 & 0 & i \sqrt{D_{-\sigma}} e^{-i\varphi (t)/2}  & \sqrt{R_{-\sigma}}  \\
\end{pmatrix}
e^{-i\theta_{\sigma}/2}.
\label{Ssfsfuntild}
\end{equation}
\end{widetext}
This matrix obeys the symmetry relation
\begin{equation}
\tilde{\mathbb{S}}_n = \hat o_1 \hat \tau_1 \hat \sigma_2 \tilde{\mathbb{S}}_n^+ \hat \sigma_2 \hat \tau_1 \hat o_1.
\label{Ssym}
\end{equation}

After the unitary transformation the operator $\hat o_1 \hat \sigma_1$ turns into a diagonal matrix 
\begin{equation}
U \hat o_1 \hat \sigma_1 U^+
=
\begin{pmatrix}
1 & 0 & 0 & 0 & 0 & 0 & 0 & 0 \\
0 & 1 & 0 & 0 & 0 & 0 & 0 & 0 \\
0 & 0 & 1 & 0 & 0 & 0 & 0 & 0 \\
0 & 0 & 0 & 1 & 0 & 0 & 0 & 0 \\
0 & 0 & 0 & 0 &-1 & 0 & 0 & 0 \\
0 & 0 & 0 & 0 & 0 &-1 & 0 & 0 \\
0 & 0 & 0 & 0 & 0 & 0 &-1 & 0 \\
0 & 0 & 0 & 0 & 0 & 0 & 0 &-1 \\
\end{pmatrix},
\end{equation}
and the matrix $\tilde W$ splits into blocks \eqref{blocks} which transform into each other under the Hermitian conjugation
\begin{equation}
W_{++} = W_{++}^+ , 
\quad W_{--} = W_{--}^+ ,
\quad W_{+-} = -W_{-+}^+ .
\end{equation}
Making use of the unitarity condition for the matrix $\tilde W$ we obtain the following relations
\begin{gather}
W_{++}^2  - W_{+-} W_{-+} =1, \quad 
W_{--}^2  - W_{-+} W_{+-} =1,
\label{Wrel1}
\\
W_{++} W_{+-} = W_{+-} W_{--}, \quad 
W_{-+} W_{++} = W_{--} W_{-+}.
\label{Wrel2}
\end{gather}
Owing to additional symmetry relations \eqref{sympm} 
it suffices to specify only the two matrices $W_{++}$ and $W_{+-}$. They have the form
\begin{gather}
W_{++}=
\begin{pmatrix}
0 & 0 &  r_+^* & -i d_{1+}^*\\
0 & 0  &  -i d_{2+}^* & r_+^*\\
r_+ & i d_{2+} &  0 & 0\\
i d_{1+} & r_+ &  0 & 0\\
\label{B8}
\end{pmatrix},
\\
W_{+-}=
\begin{pmatrix}
0 & 0 & r_-^* & -i d_{1-}^* \\
0 & 0 & -i d_{2-}^*  & r_-^* \\
-r_- & -i d_{2-} & 0 & 0 \\
-i d_{1-} & -r_- & 0 & 0 \\
\end{pmatrix},
\end{gather}
where we defined 
\begin{gather}
r_+ = \dfrac{\sqrt{R_+} e^{i\theta/2} + \sqrt{R_-} e^{-i\theta/2}}{2},
\\ 
r_- = \dfrac{\sqrt{R_+} e^{i\theta/2} - \sqrt{R_-} e^{-i\theta/2}}{2},
\\
d_{1+} = \dfrac{\sqrt{D_+}e^{i\theta/2} e^{i\varphi (t)/2} + \sqrt{D_-} e^{-i\theta/2} e^{-i\varphi (t)/2} }{2},
\\
d_{2+} = \dfrac{\sqrt{D_+}e^{i\theta/2} e^{-i\varphi (t)/2} + \sqrt{D_-} e^{-i\theta/2} e^{i\varphi (t)/2} }{2},
\\
d_{1-} = \dfrac{\sqrt{D_+} e^{i\theta/2} e^{i\varphi (t)/2} - \sqrt{D_-} e^{-i\theta/2} e^{-i\varphi (t)/2} }{2},
\\
d_{2-} = \dfrac{\sqrt{D_+}e^{i\theta/2} e^{-i\varphi (t)/2} - \sqrt{D_-} e^{-i\theta/2} e^{i\varphi (t)/2} }{2}.
\label{B15}
\end{gather}


\begin{thebibliography}{99}
\bibitem{FN} Subgap charge transfer in junctions between conventional superconductors is associated with Andreev reflection every act of which corresponds to transferring one Cooper pair with charge $2e$.
\bibitem{FNR} Here we do not consider the issue of effective charge fractionalization in interacting systems discussed in the literature in various physical contexts, cf., e.g., Refs. \onlinecite{H,RB,L,R}. This issue is not relevant for the problem in question, see, e.g., Refs. \onlinecite{SZ90,book} for an extensive discussion of the allowed charge states and their symmetries in superconducting junctions.
\bibitem{FN2} Within our approach we do not explicitly determine a numerical prefactor of order one in front of the last term in Eq.
\eqref{IVcurve} which -- according to \cite{GZPisma,GZ24} -- equals to $\simeq 0.59$. Within the same accuracy for $\theta =0$ our result \eqref{TVconsttrans3} should match with the analogous result \cite{GZPisma} for $I(t)$ provided one sets $l_{\rm max} \sim |\Delta/(eV)|^{1/3}$ in Eq. (16) of that work.
\end{thebibliography}
\end{document}